\documentclass[a4paper,11pt]{article}
\usepackage{jheppub} 
\usepackage{lineno}

\arxivnumber{to be updated} 

\title{\boldmath Global observables and identified-hadron production in pp, O–O and Pb–Pb collisions at LHC Run 3 energies with EPOS4}

 \author[a,1]{Hirak Kumar Koley\note{Corresponding author.}}
 \author[a,b]{and Mitali Mondal}
 \affiliation[a]{Nuclear and Particle Physics Research Centre, Department of Physics, Jadavpur University, Kolkata - 700032, India}
\affiliation[b]{School of Studies in Environmental Radiation and Archaeological Sciences, Jadavpur University, Kolkata - 700032, India}

\emailAdd{hirak.koley@gmail.com}
\emailAdd{mitalimon@gmail.com}

\abstract{
The observation of collectivity in small and large collision systems challenges our understanding of thermalization and particle production. EPOS4 models this via a dynamical core--corona separation, where high-density regions form a collectively expanding core while low-density regions hadronize via string fragmentation. Its microcanonical core hadronization improves the description of transverse momentum and multiplicity-dependent observables.
We present EPOS4 predictions for pp, O–O and Pb–Pb collisions, with and without UrQMD, showing non-universal $\langle p_T\rangle$ scaling, significant hadronic-phase effects, and system-size-dependent $R_{AA}$ suppression.
Charged-particle and transverse-energy densities show participant scaling; the transverse energy per charged particle is systematically larger in O--O than in Pb--Pb at comparable participant fraction, indicating a harder effective production in the lighter system. Identified-hadron spectra harden with event multiplicity with mass ordering and increasing core fractions. The mean transverse momentum exhibits a strong system dependence, with the steepest multiplicity evolution in pp, demonstrating that $\langle p_T\rangle$ does not follow universal multiplicity scaling. The $p/\pi$ ratio shows an enhanced intermediate-$p_T$ region; the suppression of the integrated $p/\pi$ at the highest Pb--Pb multiplicities is reproduced only with UrQMD, highlighting hadronic-phase effects. The nuclear modification factor shows sizeable suppression in Pb--Pb and substantial suppression in central O--O collisions. Blast-wave fits exhibit the anti-correlation between $T_{\rm kin}$ and $\langle\beta_T\rangle$, with UrQMD shifting the parameters towards lower $T_{\rm kin}$ and higher $\langle\beta_T\rangle$. These results provide a timely baseline for Run~3 measurements and for constraining the onset of medium-like effects across system size.
}

\begin{document}
\maketitle
\flushbottom

\section{Introduction}\label{sec1}

The pursuit of understanding the fundamental properties of the Quark-Gluon Plasma (QGP), the deconfined state of quarks and gluons predicted by Quantum Chromodynamics (QCD) \cite{Aoki_2006}, is the central objective of modern nuclear physics research. While heavy-ion collisions at the Relativistic Heavy-Ion Collider (RHIC) at Brookhaven National Laboratory (BNL) and the Large Hadron Collider (LHC) at CERN have firmly established the QGP as a near-perfect fluid characterized by an extraordinarily low shear viscosity-to-entropy ratio ($\eta$/s) \cite{PhysRevLett.94.111601}, the current focus has shifted toward exploring the minimal requirements for this deconfined state to form. The critical challenge stems from the unexpected discovery of collective phenomena—such as mass-dependent hardening of $p_T$ spectra and non-zero anisotropic flow ($v_n$)—in high-multiplicity proton-proton (pp) and proton-nucleus (p-A) collisions \cite{2010, 201425}, systems previously considered too small to thermalize. This puzzle necessitates a deeper theoretical understanding of how collectivity emerges, raising the core question: Is the observed flow a signature of genuine QGP formation in tiny droplets, or the manifestation of novel initial-state or non-equilibrium dynamics \cite{Nagle_2018}?

This puzzle is critically addressed by examining other definitive QGP signatures. The strong suppression of high-$p_T$ hadron yields, known as jet quenching—the signature of partonic energy loss—was initially absent in p-A systems, leading to arguments against QGP formation in small systems \cite{PhysRevLett.110.082302}. However, recent analyses from RHIC's PHENIX Collaboration, utilizing direct photons as a calibrated measure of collision centrality, have yielded fresh evidence of jet suppression in the most central d-Au collisions, strengthening the hydrodynamic interpretation \cite{Abdulameer_2025}. This crucial result—suggesting that minute, short-lived QGP droplets can be created—demands that theoretical models simultaneously account for both bulk collective flow (low $p_T$) and medium-induced energy loss (high $p_T$).

The current phase of investigation involves the introduction of collisions with light ions, such as the recently collected oxygen-oxygen (O-O) collisions at the LHC at $\sqrt{s_{\rm NN}} =$5.36 TeV \cite{citron2019futurephysicsopportunitieshighdensity}. This light-ion system serves as a crucial bridge, filling the geometric and multiplicity gap between pp/p-Pb and Pb-Pb systems \cite{Schenke_2020}. 
The core purpose of the O-O run is to provide an unparalleled environment for a controlled system size scan, which is essential for decoupling the influence of initial-state nuclear geometry from final-state medium dynamics \cite{Huang_2020}. 
Observing how signatures of collectivity—such as the scaling of anisotropic flow ($v_n$) and mass-ordering patterns in $p_T$ spectra—evolve continuously from the fragmentation-dominated pp baseline, through the newly explored O-O system, and into the hydrodynamics-dominated Pb-Pb environment is key to defining the minimal conditions required for thermalization. 
Specifically, comparing systems with similar charged-particle multiplicity density ($<\frac{dN_{ch}}{d\eta}>$) but vastly different initial geometries (e.g., pp vs. O-O vs. Pb-Pb) is therefore essential to rigorously isolate whether the observed features are governed by initial-state correlations or by late-stage collective expansion. 
This systematic geometric control allows theoretical frameworks to be stringently tested on their ability to predict the onset and nature of QGP-like phenomena as a function of volume.

Standard QCD-based event generators (e.g., PYTHIA) \cite{Sj_strand_2015}, which rely on perturbative Quantum Chromodynamics ($\text{pQCD}$) scattering and independent Lund string fragmentation \cite{ANDERSSON198331}, inherently struggle to explain the smooth, collective transition observed across systems. These models lack the dynamical mechanism necessary to transition from independent particle production to a collectively expanding fluid. While phenomenological modifications like enhanced color reconnection can mimic aspects of flow \cite{Christiansen_2015}, they do not provide a unified, a priori description of the entire system evolution. This inadequacy highlights the need for models based on dynamical medium formation, such as the EPOS framework \cite{Werner_2006}.

The EPOS model offers a realistic theoretical solution through its core-corona approach. This approach provides a unified description by dynamically separating the collision into a high-density, thermalized core (responsible for collective flow) and a fragmentation-based corona (modeling the non-collective baseline) \cite{PhysRevLett.98.152301}. This dual mechanism allows the model to naturally interpolate between systems of drastically different sizes, offering a unified explanation for the transition from pp to Pb-Pb. This study utilizes the latest iteration, EPOS4 \cite{PhysRevC.108.064903, PhysRevC.108.034904, PhysRevC.109.014910, PhysRevC.109.034918}, which incorporates critical advancements, including refined initial-state fluctuations, improved saturation and screening treatments, microcanonical statistical hadronization, and the essential coupling to the UrQMD hadronic afterburner to accurately model the non-equilibrium hadronic cascade phase \cite{Bass_1998}.

In this work, we present a comprehensive study using the EPOS4 framework to investigate identified hadron production across a wide range of collision systems: pp at $\sqrt{s} =$13.6 TeV, and O-O and Pb-Pb at $\sqrt{s_{\rm NN}} =$5.36 TeV. 
We calculate key observables for charged particles—including pseudorapidity ($\eta$) and multiplicity distributions, and transverse energy density (d$E_T$/d$\eta$)—and for identified hadrons—including $p_T$ spectra, integrated yields, particle ratios, and the nuclear modification factor ($R_{AA}$), all aiming to provide a unified description and robust predictions for the full range of LHC Run 3 systems.
A central objective is to leverage the EPOS4 core-corona approach to dissect the origins of collectivity by quantifying the relative contributions of the hydrodynamic core across varying system sizes and multiplicities. Furthermore, we investigate kinetic freeze-out dynamics using the Blast-Wave model fit to the identified hadron spectra \cite{PhysRevC.48.2462}. This analysis will provide crucial insights into the role of both initial geometry and final-state dynamics, contributing significantly to the effort to establish the definitive minimum volume or energy density threshold required for the formation of the QGP. 

The paper is organized as follows: Section II details the theoretical features and recent upgrades of EPOS4. Section III presents the results for the predicted observables and comparisons with experimental data. Section IV concludes with a summary of the key findings and a discussion of their implications for the system-size dependence of collective dynamics.

\section{The EPOS4 Model}\label{sec2}

The EPOS4 framework is the latest iteration in the EPOS series, designed as a full-chain Monte Carlo event generator to provide a unified and dynamic description of particle production across all collision systems, from proton-proton ($\text{pp}$) to nucleus-nucleus ($\text{A-A}$). The model is fundamentally rooted in a pQCD inspired Gribov-Regge multiple scattering approach \cite{DRESCHER200193}. A significant theoretical advancement in EPOS4 is the implementation of a rigorously parallel scattering scheme for primary interactions, replacing the previous sequential treatment \cite{PhysRevC.109.034918}. This new scheme ensures energy-momentum conservation is respected across all parallel sub-scatterings, which is necessary for consistency at ultra-relativistic energies and for accurately describing event-by-event fluctuations in the initial state.

The core of the model's unifying power is the core-corona approach, a dynamic separation mechanism based on local energy density. Initial interactions produce numerous parton-based flux tubes (strings), or pre-hadrons. These strings are dynamically separated: those exceeding a model-dependent energy loss or density threshold form the dense core, while the remaining dilute segments form the corona. The core matter is assumed to be thermalized and is mapped onto an energy-momentum tensor, which is then subjected to a full $3+1$D viscous hydrodynamic evolution \cite{Werner_2010}. This module utilizes full 3D initial conditions that include event-by-event fluctuations in space and time, allowing for a more realistic modeling of the anisotropic fireball structure.

The hydrodynamic evolution is governed by the equation of state and the transport properties of the medium. A key transport parameter used is the shear viscosity-to-entropy ratio ($\eta/s$) \cite{PhysRevLett.94.111601}. This ratio quantifies the fluidity of the matter; a lower value indicates a more perfect fluid, aligning with the properties established for the QGP. In EPOS4, $\eta/s$ is typically set to a low value (e.g., $\eta/s = 0.08$) to model the near-perfect fluid behavior of the core. The core is thus the source of collective flow and QGP-like phenomena, while the corona segments hadronize independently via standard string fragmentation processes, defining the non-collective baseline.

The hydrodynamic evolution continues until the system reaches a defined hadronization hypersurface ($\epsilon_H = 0.57$ $\text{GeV}/\text{fm}^3$) \cite{PhysRevC.108.064903}. Hadronization occurs via a microcanonical statistical approach on this surface, which ensures the exact conservation of conserved quantities (baryon number, charge, strangeness). This microcanonical treatment is a crucial advancement for small systems, as it naturally accounts for the strangeness suppression observed in low-multiplicity events where the available phase space for statistical sampling is restricted. Finally, all final-state hadrons are passed to the Ultra-relativistic Quantum Molecular Dynamics (UrQMD) afterburner \cite{Bass_1998} to simulate the subsequent non-equilibrium hadronic cascade phase—the final elastic and inelastic re-scattering and resonance decay—which determines the evolution toward kinetic freeze-out. 

\begin{figure}[t]
	\centering 
    \includegraphics[width=0.49\textwidth, angle=0]{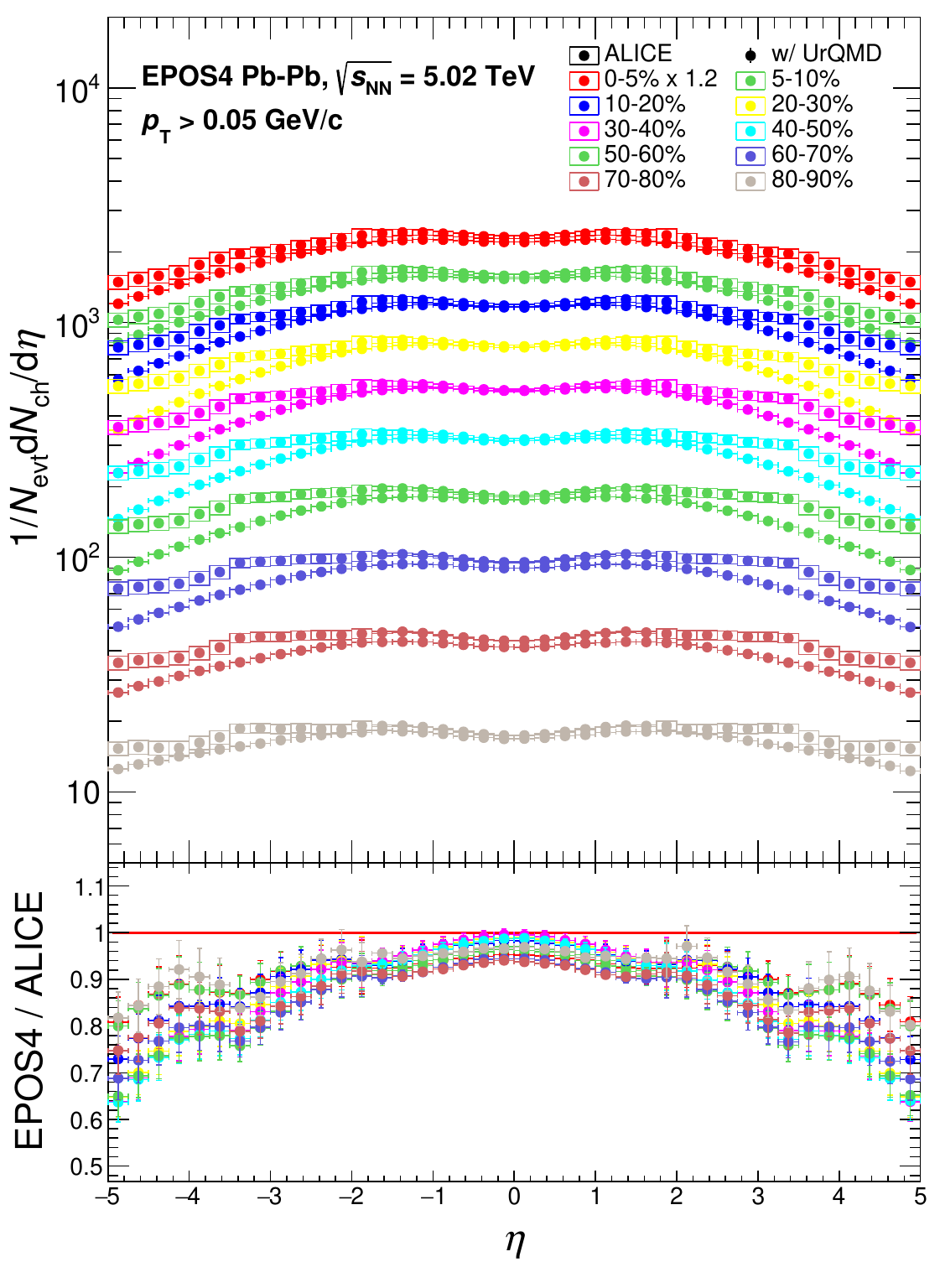}	
	\includegraphics[width=0.49\textwidth, angle=0]{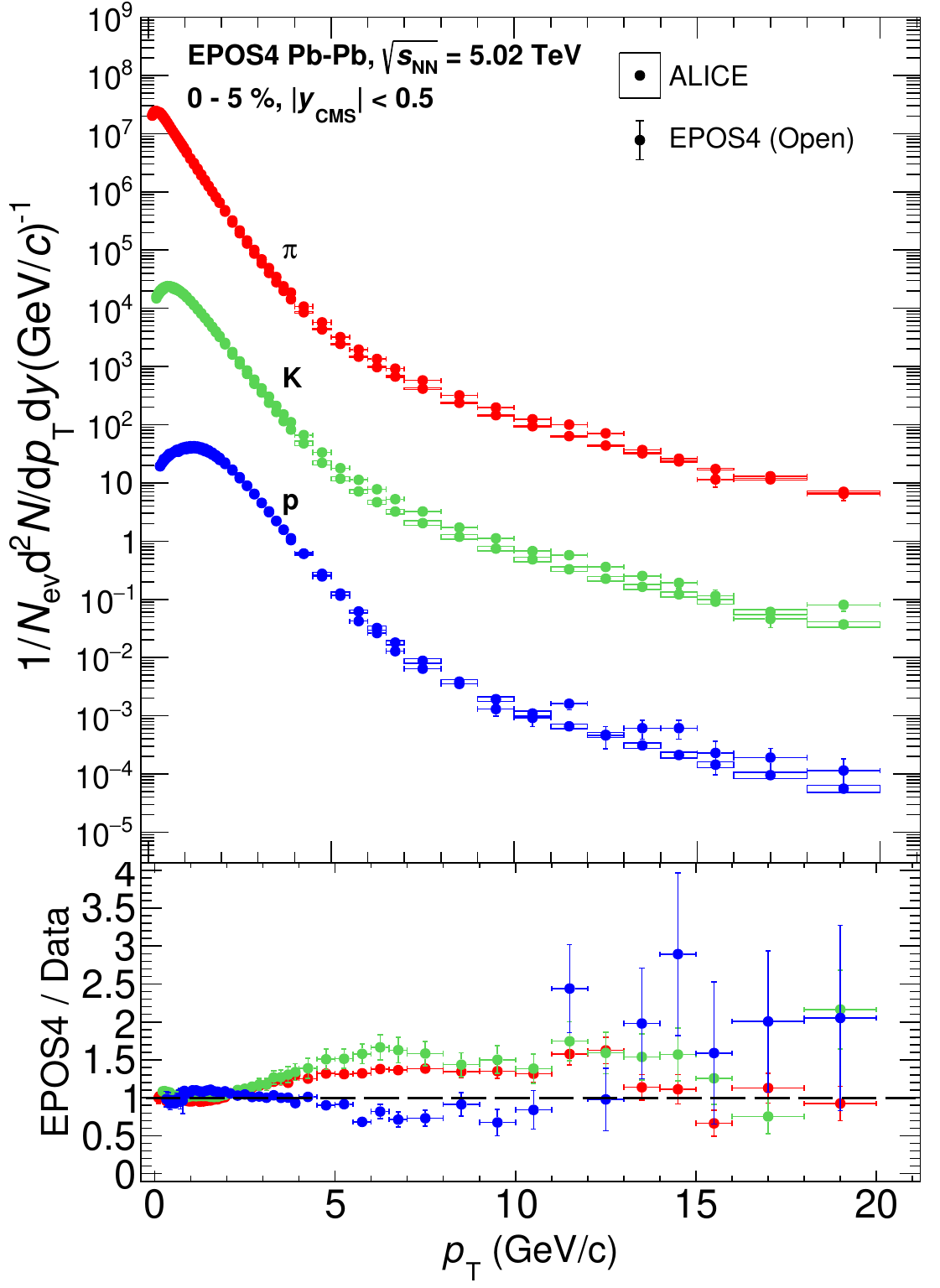}	
	\caption{(Color online) (left) Pseudorapidity density of primary charged particles for different centrality classes over a broad $\eta$ range in Pb-Pb collisions at $\sqrt{s_{NN}} = 5.02$ TeV and (right) Transverse momentum spectra of pions, kaons, and (anti-)protons in Pb-Pb collisions at $\sqrt{s_{NN}}$ = 5.02 TeV for 0-5$\%$ centrality class, compared with ALICE \cite{PhysRevC.101.044907, 2017567}.} 
	\label{EPOSvalidation}%
\end{figure}

For this study, we generated 20 million minimum-bias $\text{pp}$ events at $\sqrt{s} = 5.02$, 5.36, and 13.6 TeV; 2 million $\text{O–O}$ events at $\sqrt{s_{\mathrm{NN}}} = 5.36$ TeV; and 0.5 million $\text{Pb–Pb}$ events at $\sqrt{s_{\mathrm{NN}}} = 5.02$ and 5.36 TeV.
For each system and energy, simulations were performed both with and without the inclusion of the UrQMD hadronic afterburner to systematically quantify the influence of the final-state hadronic cascade.
The $5.02$ TeV $\text{pp}$ and $\text{Pb-Pb}$ samples were used for model validation against Run 2 data, while the high-energy $\text{pp}$, $\text{O-O}$, and $\text{Pb-Pb}$ samples form the basis for our new analysis and Run 3 predictions.
In the presentation of our results, samples incorporating the afterburner are designated as "w/ UrQMD," while those without are referred to as "w/o UrQMD." For our detailed analysis concerning the origins of collectivity, a critical methodological point is that the essential core–corona separation analysis is exclusively performed on samples w/o UrQMD. This is necessary because EPOS assigns each pre-hadron an internal tag indicating its origin (core or corona), and this unique tag is only guaranteed to be preserved before the final-state interactions modelled by UrQMD.

To explore the dependence of particle production on system size and geometry, we classified the minimum-bias event samples into several multiplicity percentile classes. This classification is based on the total charged-particle multiplicity measured in the combined ALICE V0A and V0C acceptances \cite{collaboration_2013}, specifically the $\eta$ ranges $2.8 < \eta_{\text{lab}} < 5.1$ (V0A) and $-3.7 < \eta_{\text{lab}} < -1.7$ (V0C). 
To establish the reliability of the EPOS4 model, particularly for the heavy-ion dynamics that govern the core's hydrodynamic evolution, we performed a stringent validation against existing $\text{ALICE}$ Run 2 data in $\text{Pb-Pb}$ collisions at $\sqrt{s_{\text{NN}}} = 5.02$ TeV. The validation of the $\text{pp}$ at same energy was previously detailed in our earlier work \cite{Koley_2025}. We first compare the $\text{Pb-Pb}$ charged-particle pseudorapidity distributions ($\frac{1}{N_{\text{ev}}}\frac{\text{d}N_{\text{ch}}}{\text{d}\eta}$) across various centrality classes \cite{2017567}. The left panel of Fig. \ref{EPOSvalidation} demonstrates that the model successfully describes the complex centrality dependence and quantitatively reproduces the mid-rapidity particle density across the full range when $\text{UrQMD}$ is included, confirming the accuracy of the initial geometry and energy deposition within the model.
Furthermore, we critically compare the $\text{p}_T$-differential invariant yields for identified hadrons ($\pi$, $\text{K}$, $\text{p}$) against $\text{ALICE}$ data in the $0-5\%$ most central $\text{Pb-Pb}$ collisions \cite{PhysRevC.101.044907}. The right panel of Fig. \ref{EPOSvalidation} clearly confirms that EPOS4 captures the overall spectral shape of the $\text{p}_T$ spectra in similar kinematic ranges. The figure’s ratio panel indicates that EPOS4 provides a good quantitative description of the low-$\text{p}_T$ bulk spectra, successfully validating the model's complex initial-state, hydrodynamic expansion, and hadronization mechanisms. However, a discrepancy peaking up to approximately $40\%$ is observed at intermediate $\text{p}_T$, though the data-to-model ratios approach unity at higher-$\text{p}_T$, indicating gradual improvement in the quantitative description. Based on this robust validation in both the pp and most demanding $\text{Pb-Pb}$ systems, we conclude that the EPOS4 framework is a well-tested tool and is suitably configured to provide the detailed predictions for the LHC Run 3 data \cite{koley2026epos4modelpredictionsglobal}, particularly the novel $\text{O-O}$ system.

\section{Results}\label{sec3}

\subsection{Pseudorapidity Density of Charged Particles}\label{subsec2}
\begin{figure}[h]
	\centering 
	\includegraphics[width=1.0\textwidth, angle=0]{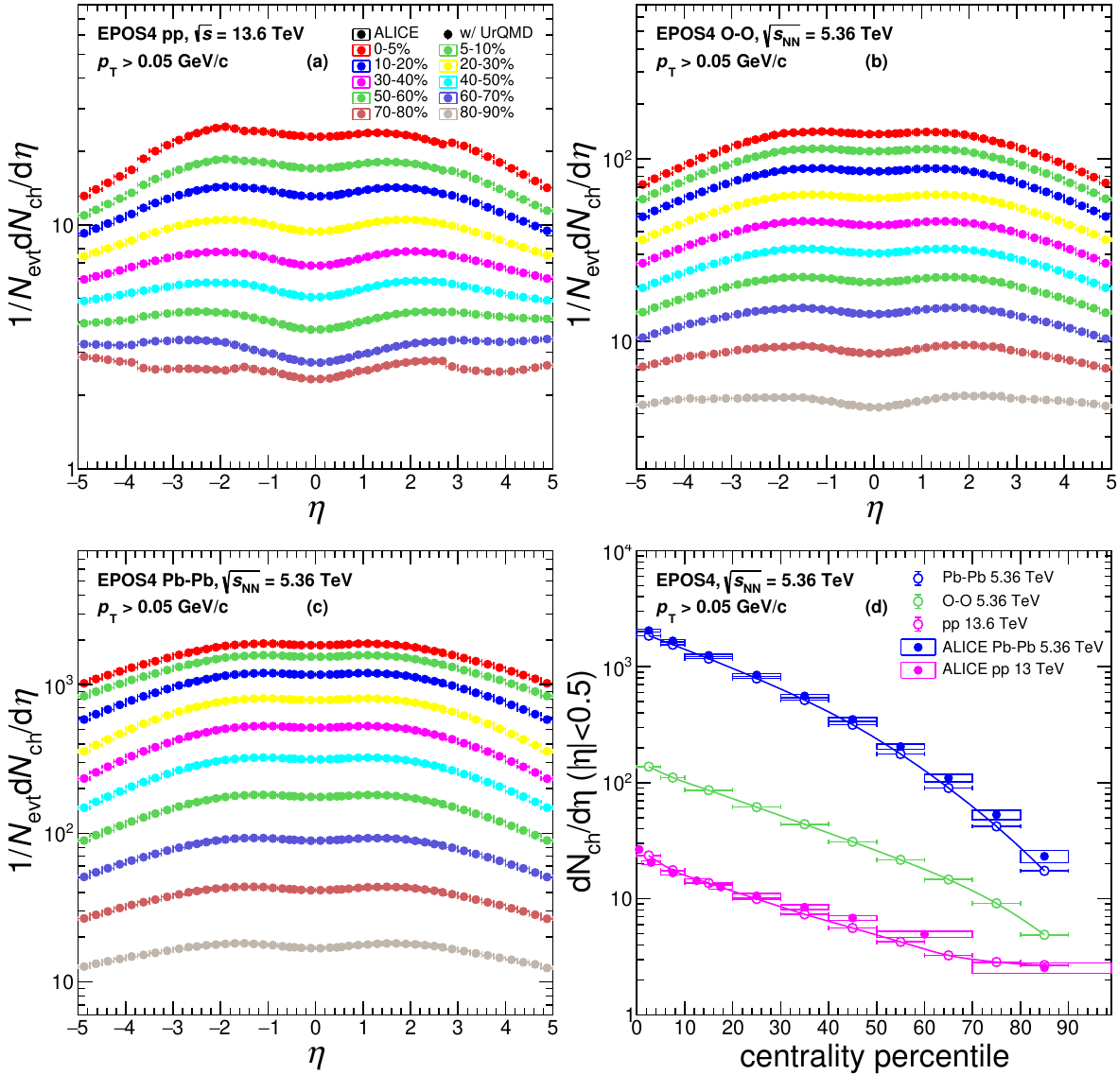}	
	\caption{(Color online) (a-c) Pseudorapidity density of primary charged particles for different centrality classes over a broad $\eta$ range in  pp collisions at $\sqrt{s} = 13.6$ TeV and O-O and Pb–Pb collisions at $\sqrt{s_{NN}}$ = 5.36 TeV, and (d) Average charged particle multiplicity as a function of centrality percentile, compared with ALICE \cite{2019, alicecollaboration2025centralitydependencechargedparticlepseudorapidity}.} 
	\label{EtaDistOO}%
\end{figure}

The pseudorapidity density of primary charged particles, $dN_{\text{ch}}/d\eta$, is a key global observable in high-energy nuclear collisions that provides direct insight into the overall particle production and initial entropy density of the system. 
Its dependence on the system size and collision centrality plays an essential role in characterizing the initial state and in constraining hydrodynamic and transport calculations.

We investigate $dN_{\text{ch}}/d\eta$ at midrapidity ($|\eta| < 0.5$) in pp collisions at $\sqrt{s} = 13.6$ TeV and in O-O and Pb–Pb collisions at $\sqrt{s_{NN}}$ = 5.36 TeV using the EPOS4 event generator.
Figures \ref{EtaDistOO} (a-c) present the pseudorapidity distributions of primary charged particles across a broad $\eta$ range for several centrality classes.
Across all systems, EPOS4 reproduces the characteristic shape: an approximately flat plateau around midrapidity followed by a gradual decline toward forward and backward pseudorapidities.
The overall particle yield increases systematically with centrality, reflecting the larger number of initial partonic scatterings and higher entropy density in more central collision geometries.
The centrality dependence of the midrapidity density, shown in Fig. \ref{EtaDistOO} (d), exhibits a smooth, monotonic rise from peripheral to central events.
In addition, the O–O midrapidity densities populate the region between pp and Pb–Pb values, following the expected system-size hierarchy and further confirming its role as an intermediate system that bridges small and large collision systems, enabling a more differential study of initial-state and final-state effects.
Comparisons with ALICE Run 2 measurements for pp collisions at $\sqrt{s}=13$ TeV \cite{2019}, together with the corresponding EPOS4 predictions at $\sqrt{s}=13.6$ TeV, demonstrate that the model provides a reasonable description of both the overall magnitude and the centrality evolution of $dN_{\text{ch}}/d\eta$.
However, modest deviations visible in peripheral Pb-Pb events at $\sqrt{s_{NN}}$ = 5.36 TeV may originate from limitations in modelling initial-state fluctuations at the higher collision energy \cite{alicecollaboration2025centralitydependencechargedparticlepseudorapidity}.
Since $dN_{\text{ch}}/d\eta$ forms the basis of centrality classification and normalizes numerous key observables—including transverse energy, identified-hadron spectra, and particle-ratio systematics—its accurate theoretical description is essential. The present results demonstrate that EPOS4 captures the essential trends of particle production across the full range of pp and Pb–Pb collisions and provides robust predictions for forthcoming Run 3 measurements, particularly for the recently recorded O–O collision system at the LHC.

\begin{figure}[h]
	\centering 
	\includegraphics[width=0.48\textwidth, angle=0]{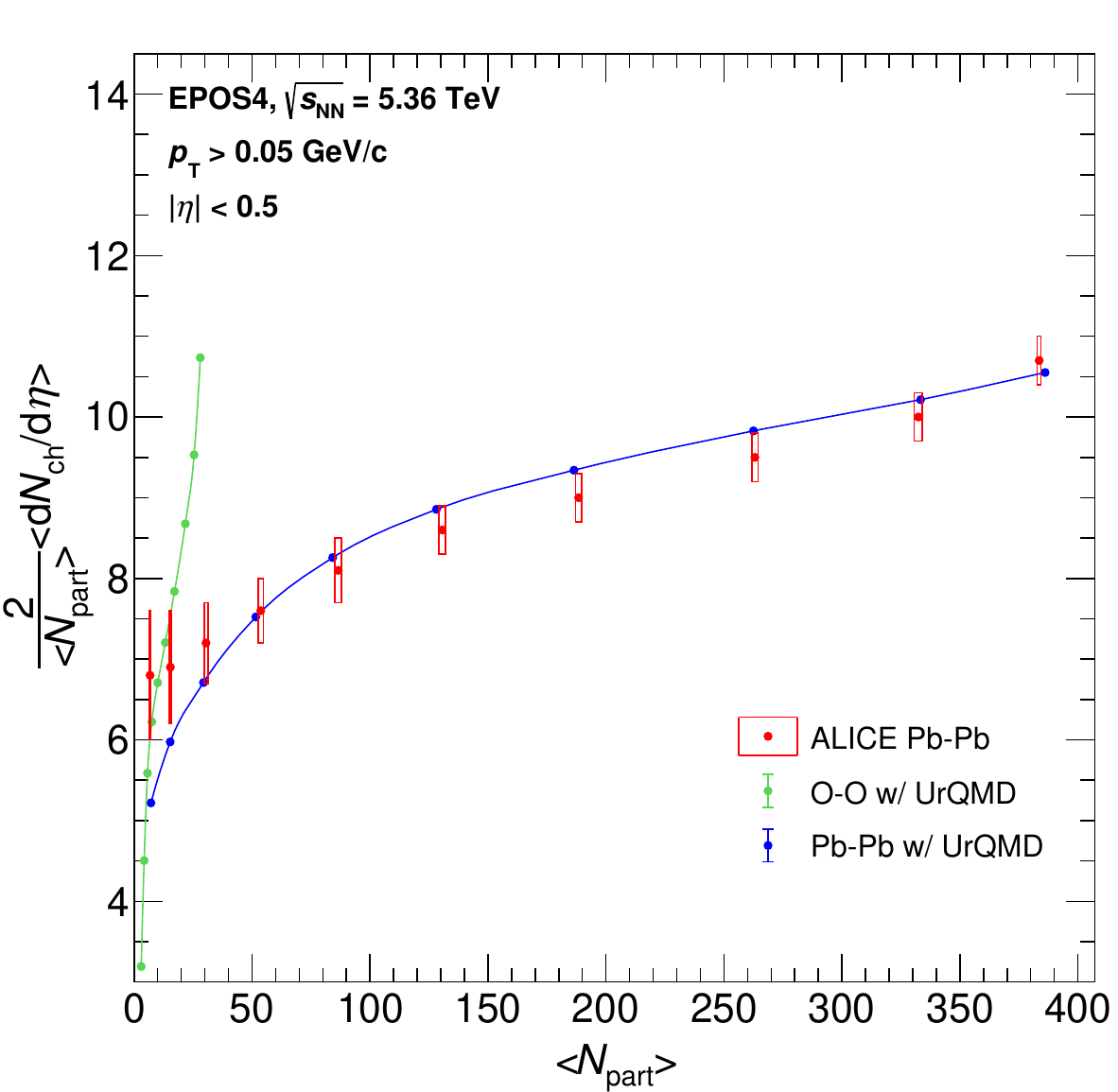}	
    	\includegraphics[width=0.48\textwidth, angle=0]{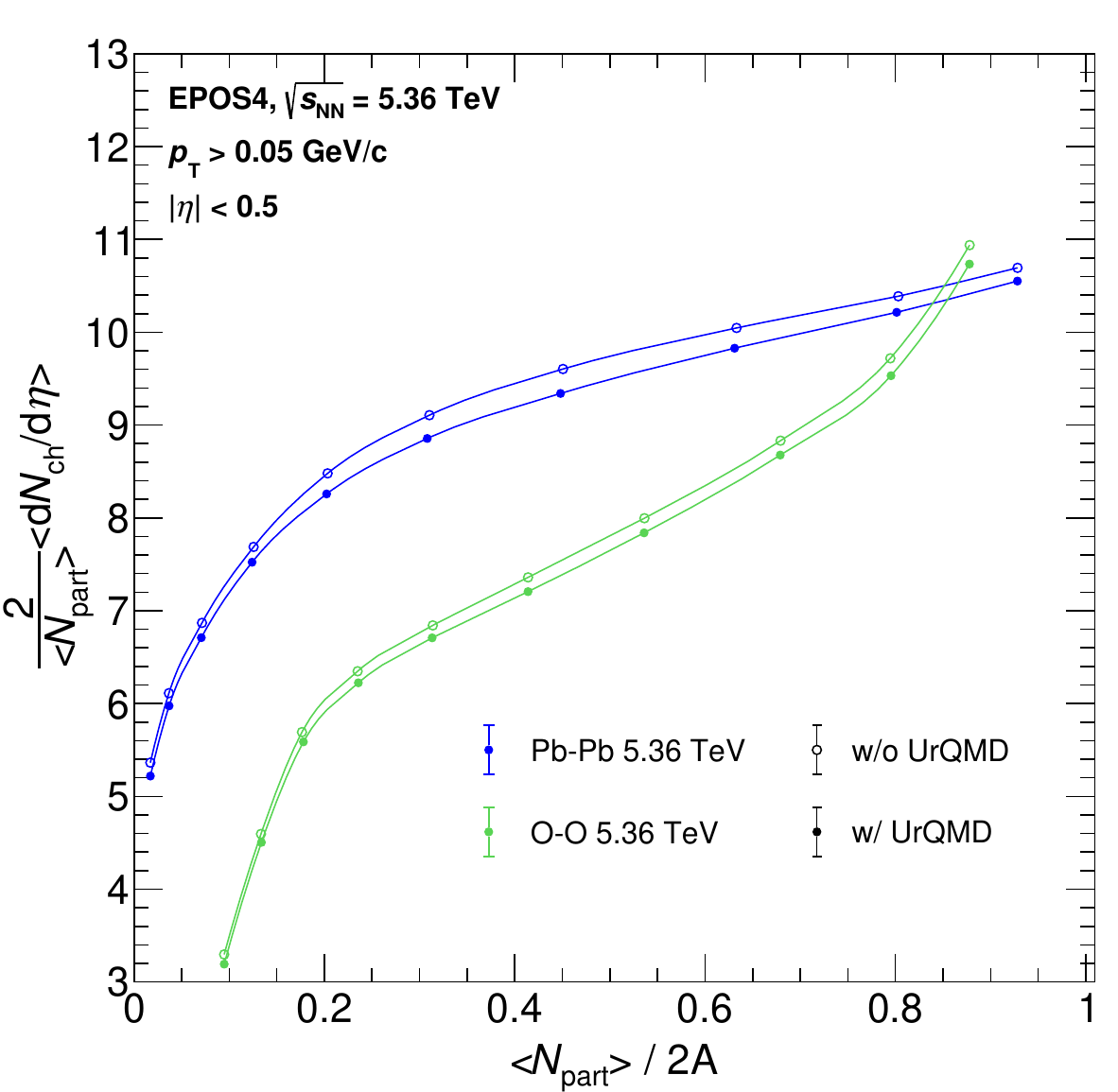}	
	\caption{(Color online) (Left) Charged-particle multiplicity density at midrapidity, $\frac{dN_{ch}}{d\eta}$, as a function of the number of participating nucleons, $\langle N_{\text{part}}\rangle$, for O–O and Pb–Pb collisions at $\sqrt{s_{NN}}$ = 5.36 TeV, compared with ALICE data \cite{alicecollaboration2025centralitydependencechargedparticlepseudorapidity} and (Right) The same observable shown as a function of the scaled participant fraction, $\langle N_{\text{part}}\rangle/2A$.} 
	\label{Npart1}%
\end{figure}

\subsection{Charged Particle Multiplicity Density}\label{subsubsec2}

The average charged-particle multiplicity density at midrapidity, $\langle dN_{\text{ch}}/d\eta\rangle$, is a key global observable that reflects the overall particle production and is directly correlated with the initial entropy and energy density of the system. Its scaling with the number of participating nucleons provides a robust means of characterizing the collision geometry, validating initial-state modelling, and establishing centrality classes for differential studies \cite{alicecollaboration2025centralitydependencechargedparticlepseudorapidity, Acharya_2019}.

Figure~\ref{Npart1} (left) shows $\langle dN_{\text{ch}}/d\eta\rangle$ at \(|\eta| < 0.5\) as a function of the average number of participating nucleons, $\langle N_{\text{part}}\rangle$, for O-O and Pb–Pb collisions at \(\sqrt{s_{NN}} =\) 5.36 TeV obtained from EPOS4 with the UrQMD hadronic afterburner. 
The model provides an excellent description of the ALICE Pb–Pb data over most of the centrality range, reproducing both the magnitude and the expected monotonic increase of particle production with geometric overlap \cite{alicecollaboration2025centralitydependencechargedparticlepseudorapidity}.
A small deviation is visible at the lowest $\langle N_{\text{part}}\rangle$, where EPOS4 slightly underestimates the multiplicity.
This discrepancy is most likely related to modelling limitations in peripheral events rather than a genuine physical fluctuation.
A related behaviour was previously reported by ALICE in Xe-Xe collisions~\cite{Acharya_2019}, where a steeper-than-expected rise toward central events was attributed to event-by-event multiplicity (volume) fluctuations in the high-multiplicity tail and to finite-resolution effects of the centrality estimator.
A comparable steepening is observed for O–O collisions at large $\langle N_{\text{part}}\rangle$, naturally amplified in lighter systems where a small number of participants enhances the relative impact of multiplicity fluctuations and geometric biases stemming from the compact nuclear overlap \cite{loizides2025glauberpredictionsoxygenneon}.

The corresponding results as a function of the scaled participant fraction, 
$\langle N_{\text{part}}\rangle/2A$, are displayed in Fig.~\ref{Npart1} (right), where EPOS4 calculations with and without the UrQMD afterburner are compared.
The modest separation between the two curves indicates that hadronic re-scattering has only a limited effect on the final charged-particle multiplicity density, though small shifts in O–O, especially toward central events, indicate that the hadronic re-scattering contributes modestly but non-negligibly even in intermediate-sized systems.
The observation that O–O approaches, and in scaled form may slightly exceed, the Pb–Pb trend at large participant fractions is consistent with the enhanced sensitivity of small nuclei to multiplicity fluctuations and does not reflect an anomalously large entropy deposition.

\begin{figure}[t]
	\centering 
    \includegraphics[width=0.49\textwidth, angle=0]{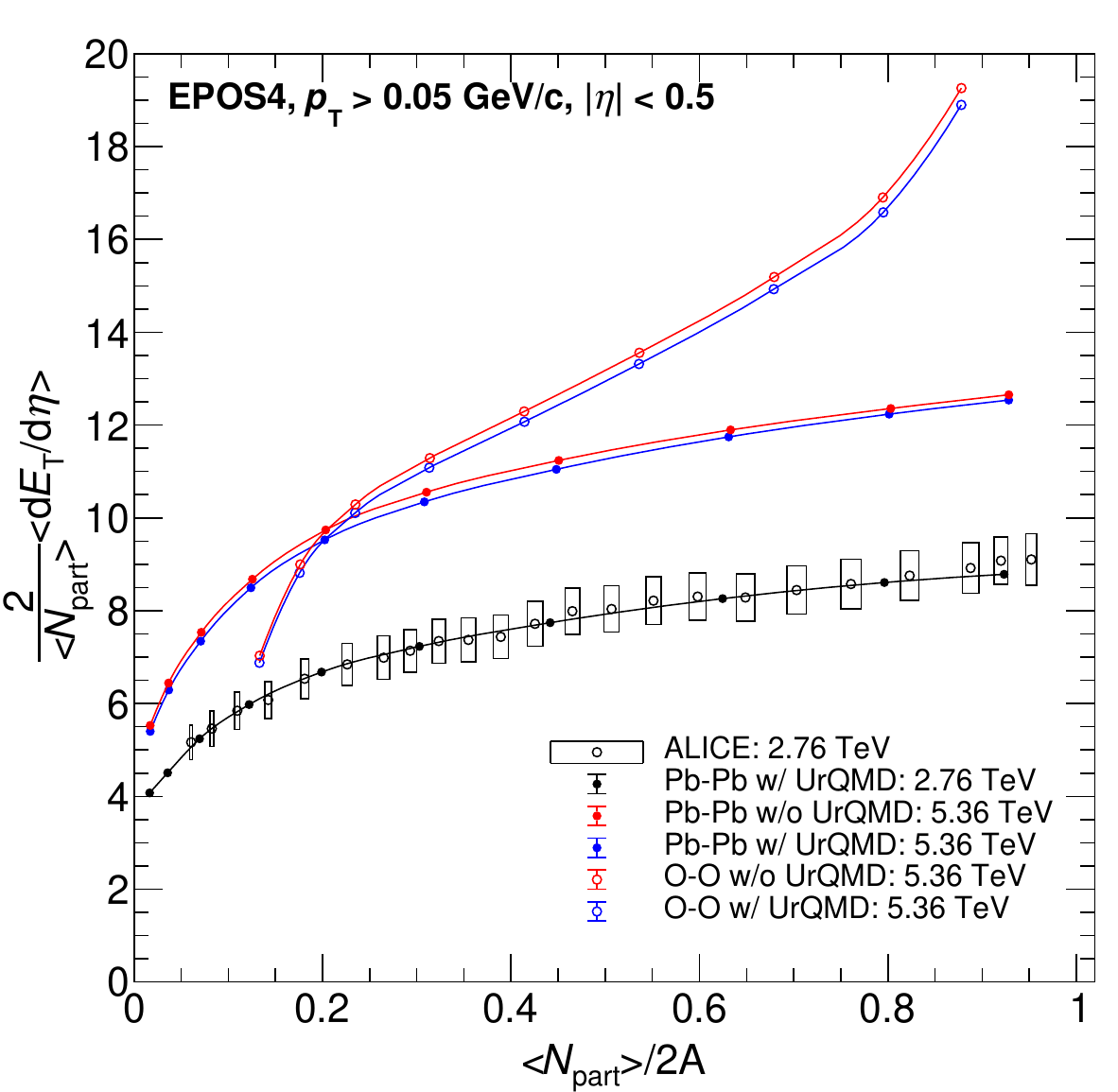}
	\includegraphics[width=0.49\textwidth, angle=0]{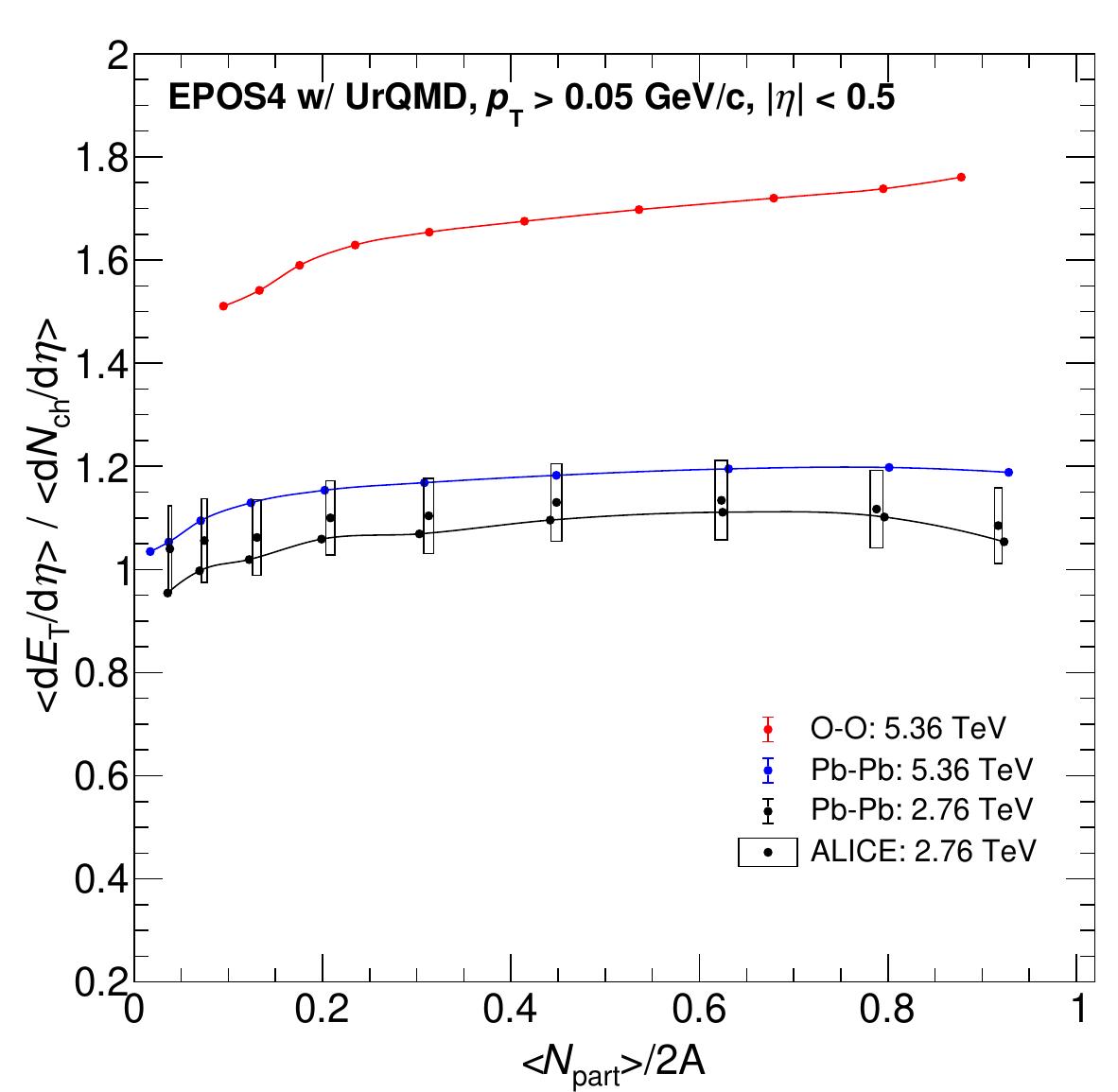}	
	\caption{(Color online) (left) Midrapidity transverse energy density $\langle dE_T/d\eta\rangle$, and (right) Transverse energy per charged particle, $\langle dE_T/d\eta\rangle$)/($\langle dN_{ch}/d\eta\rangle$, shown as a function of the scaled average participant fraction, $\langle N_{\text{part}}\rangle$/2A, for O-O and Pb-Pb collisions at $\sqrt{s_{NN}}$ = 5.36 TeV. Results are compared to available ALICE measurements \cite{PhysRevC.94.034903}.} 
	\label{Et2}%
\end{figure}

\subsection{Transverse Energy Density}\label{sec4}

The transverse energy density, $\langle dE_T/d\eta\rangle$, provides direct information on the amount of energy deposited per unit pseudorapidity in a collision and is closely related to the initial energy density and entropy production of the system \cite{SIMAK1988159}. When studied together with the charged-particle multiplicity density, it offers additional insight into the hardness of particle production and the relative contributions of soft and hard processes across different collision systems.

Experimentally, the transverse energy is defined as the sum of the transverse components of the final-state particle energies,
\begin{equation}
E_T = \sum_i E_i \sin\theta_i ,
\end{equation}
where the sum runs over all particles within the pseudorapidity interval $|\eta|<0.5$, $E_i$ is the particle energy, and $\theta_i$ is the polar angle with respect to the beam axis. In this analysis, the transverse energy is evaluated using the exact kinematic relation
$E_i \sin\theta_i = m_{T,i}$, such that
\begin{equation}
E_T = \sum_i m_{T,i},
\end{equation}
with $m_{T,i}=\sqrt{p_{T,i}^2+m_i^2}$ denoting the transverse mass of particle $i$.
Accordingly, the transverse energy density $\langle dE_T/d\eta\rangle$ corresponds to the sum of transverse masses per unit pseudorapidity. The ratio ($\langle dE_T/d\eta\rangle$)/($\langle dN_{ch}/d\eta\rangle$) therefore provides a direct measure of the average transverse energy per produced charged particle and serves as a sensitive probe of the hardness of the underlying particle-production mechanisms.

The left panel of Fig.~\ref{Et2} presents transverse-energy observables in O–O and Pb–Pb collisions at $\sqrt{s_{NN}}$ = 5.36 TeV as a function of the scaled average participant fraction, $\langle N_{\text{part}}\rangle$/2A.
The left panel shows that the midrapidity transverse energy density $\langle dE_T/d\eta\rangle$ increases monotonically with $\langle N_{\text{part}}\rangle$/2A for both systems, reflecting the growing energy deposition with increasing collision overlap. 
At a fixed scaled participant fraction, the O–O values approach and can slightly exceed those of Pb–Pb collisions. This behaviour does not indicate a larger absolute energy deposition in O–O collisions; rather, it reflects the enhanced sensitivity of per-participant observables to event-by-event multiplicity fluctuations and geometric biases, which become increasingly pronounced in lighter collision systems, as already discussed in the context of the charged-particle multiplicity density.
The EPOS4 results for Pb–Pb collisions are found to be in good agreement with ALICE measurements at $\sqrt{s_{NN}}$ = 2.76 TeV, providing confidence to the model description of bulk energy deposition \cite{PhysRevC.94.034903}.

The ratio ($\langle dE_T/d\eta\rangle$)/($\langle dN_{ch}/d\eta\rangle$) as a function of $\langle N_{\text{part}}\rangle$/2A is shown in the right panel of Fig.~\ref{Et2}, which effectively represents the average transverse energy carried by each produced charged particle and is therefore directly sensitive to the average transverse mass scale of particle production. The ratio is systematically larger in O–O collisions than in Pb–Pb collisions, indicating a higher average transverse mass per particle in the lighter system. Such a hierarchy is consistent with experimental observations that smaller systems exhibit harder transverse-momentum spectra and larger mean transverse momenta, reflecting a reduced dominance of soft, thermalized bulk production and a comparatively larger contribution from hard and semi-hard processes.

Within EPOS4, this system-size dependence emerges naturally from the interplay between core and corona particle production. In O–O collisions, corona-like emission remains significant even at high multiplicities, leading to harder spectra and enhanced transverse energy per particle, whereas in Pb–Pb collisions soft hydrodynamic emission dominates. The modest differences between calculations with and without the UrQMD hadronic afterburner indicate that final-state hadronic re-scattering has only a limited impact on transverse-energy observables and does not alter the observed system-size hierarchy.

\begin{figure}[h]
	\centering 
    \includegraphics[width=1.\textwidth, angle=0]{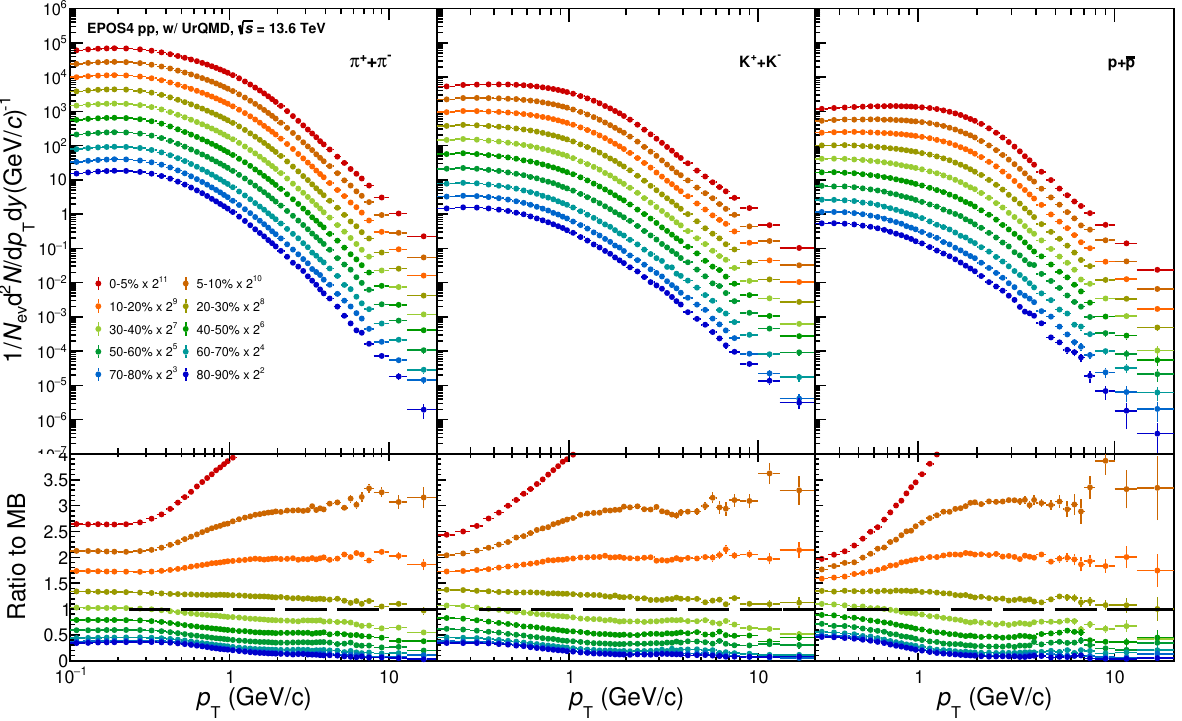}	
	\includegraphics[width=1.\textwidth, angle=0]{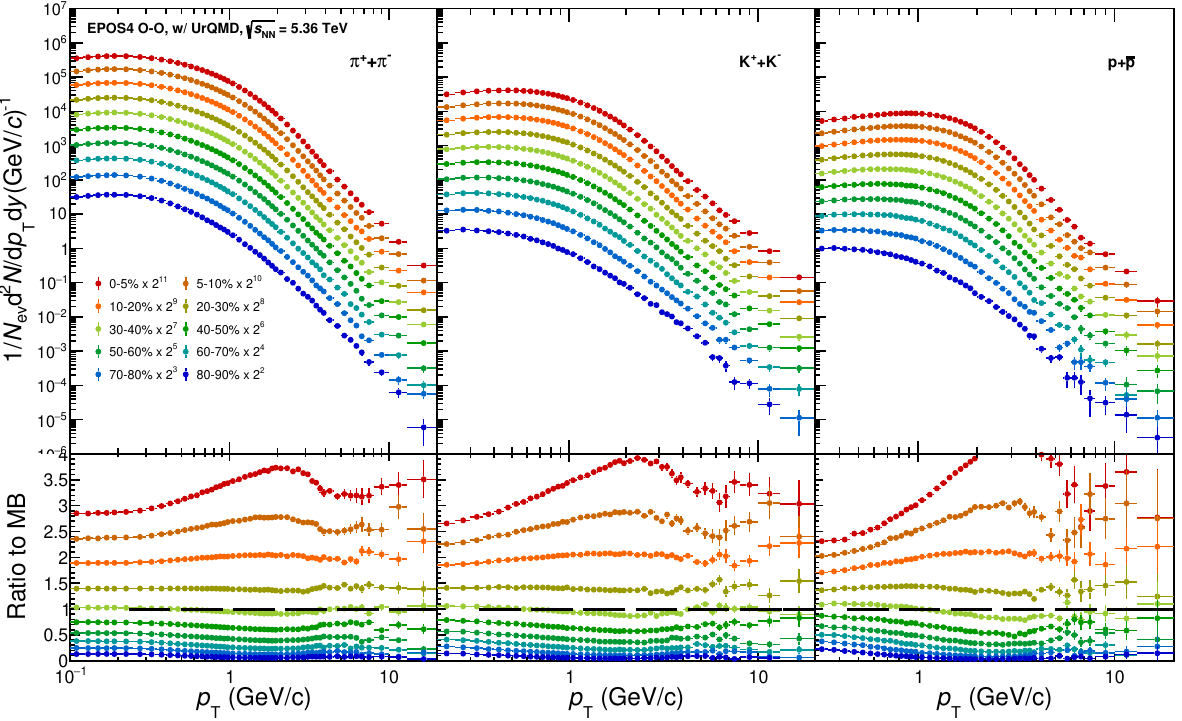}	
	\caption{(Color online) Transverse momentum spectra of identified charged hadrons $\pi^{\pm}$, K$^{\pm}$ and p($\bar{p}$) in (top) pp collisions at $\sqrt{s}$ = 13.6 TeV and in (bottom) O-O collisions at $\sqrt{s_{NN}}$ = 5.36 TeV for several multiplicity or centrality classes from EPOS4. For each systems, the upper panels show the $p_T$-differential yields, while the lower panels display the ratios of the spectra in each multiplicity class with respect to the corresponding minimum-bias reference.} 
	\label{cDrawSpectraOOpp}%
\end{figure}

\begin{figure}[h]
	\centering 	
    	\includegraphics[width=1.\textwidth, angle=0]{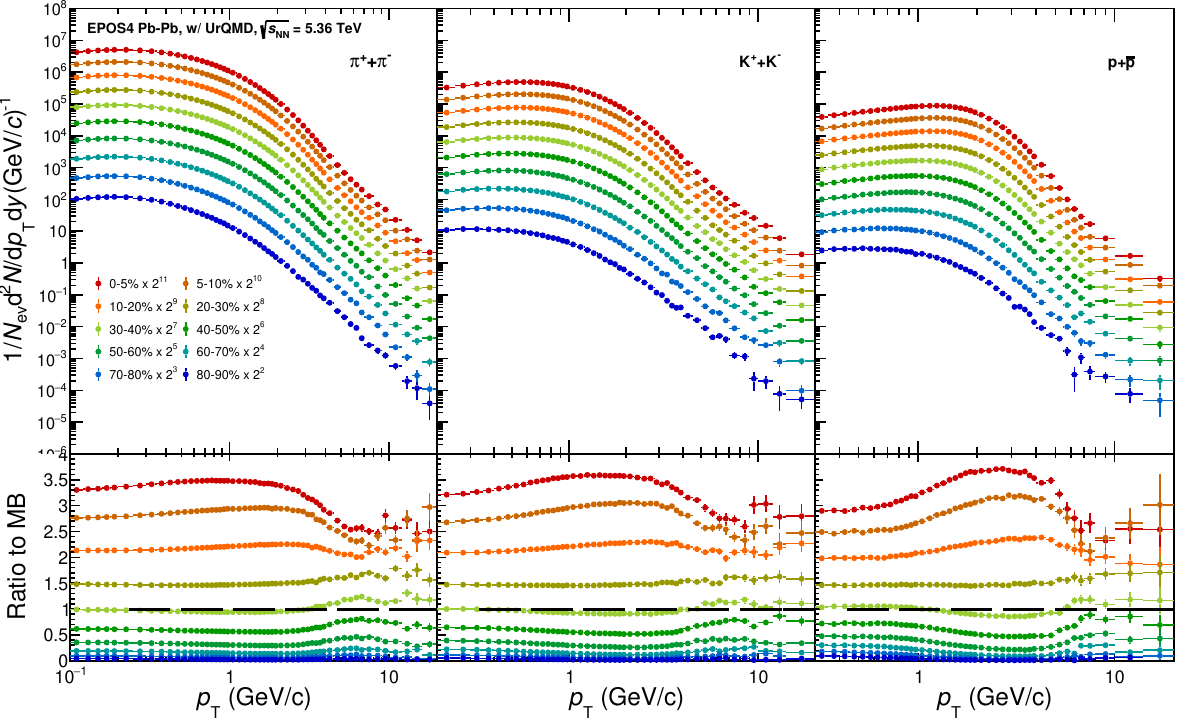}	
	\caption{(Color online) Transverse momentum spectra of identified charged hadrons $\pi^{\pm}$, K$^{\pm}$ and p($\bar{p}$) in Pb–Pb collisions at $\sqrt{s_{NN}}$ = 5.36 TeV for several centrality classes, obtained from EPOS4 simulations. The upper panels show the $p_T$-differential yields, while the lower panels display the ratios of the spectra in each centrality class to the corresponding minimum-bias reference.} 
	\label{cDrawSpectraPbPb}%
\end{figure}

\begin{figure}
	\centering 
	\includegraphics[height=0.44\textheight, angle=0]{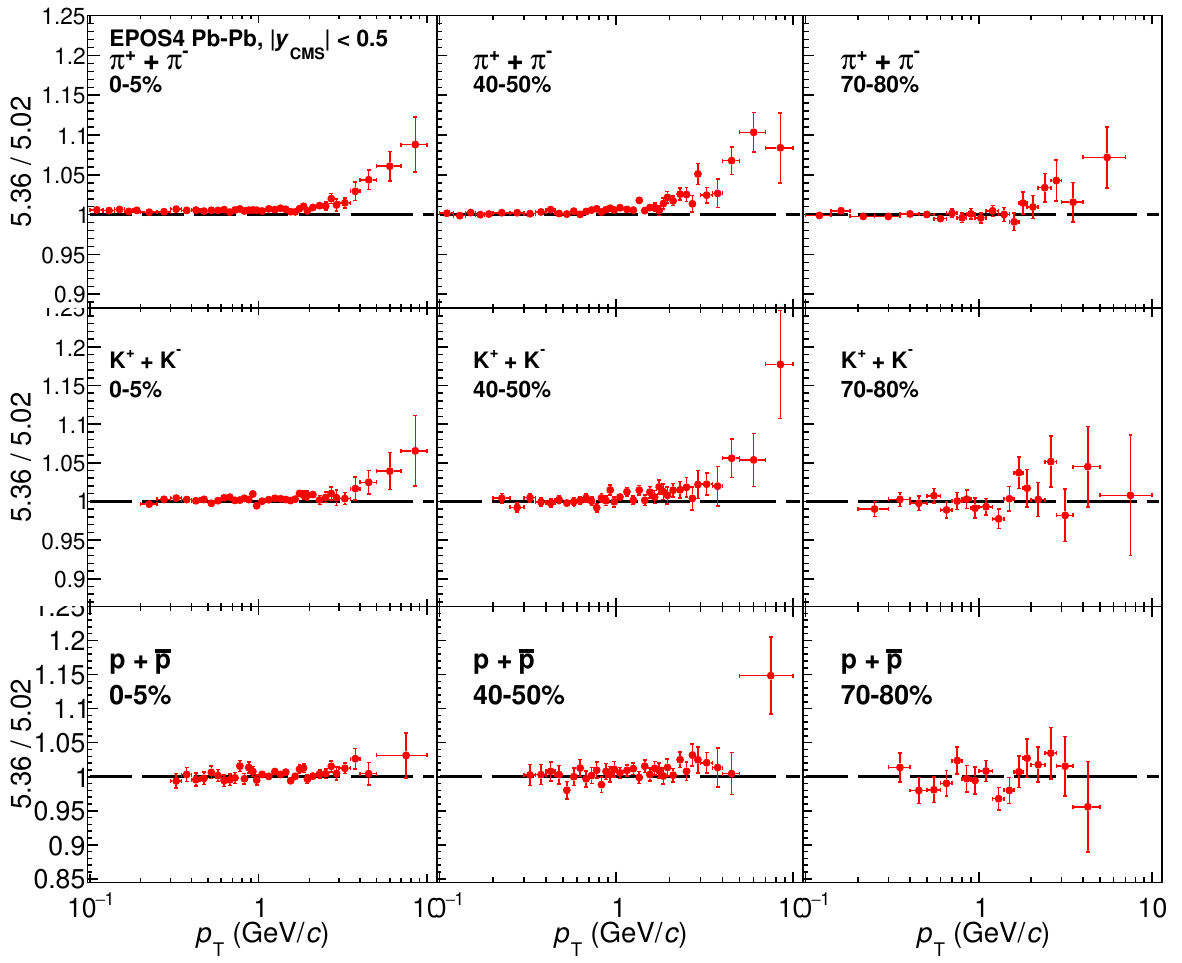}	
	\caption{(Color online) Ratios of identified-hadron $p_T$ spectra in Pb–Pb collisions at $\sqrt{s_{NN}}$ = 5.36 TeV to those at $\sqrt{s_{NN}}$ = 5.02 TeV from EPOS4 for three representative centrality classes (0–5$\%$, 40-50$\%$, and 70–80$\%$). Results are shown separately for $\pi^{\pm}$, K$^{\pm}$ and p($\bar{p}$).} 
	\label{Ratio536by502}%
\end{figure}

\subsection{Transverse Momentum Spectra and Core–Corona Composition}\label{sec5}

The transverse momentum ($p_T$) spectra of identified charged hadrons—pions ($\pi^{\pm}$), kaons (K$^{\pm}$), and (anti-)protons (p($\bar{p}$))—provide stringent constraints on the microscopic mechanisms governing particle production in relativistic nuclear collisions \cite{2017567}. At low $p_T$, the spectra are predominantly shaped by collective expansion and thermal emission, while the intermediate-$p_T$ region reflects the interplay between radial flow, hadronization dynamics, and semi-hard processes \cite{Du_2026}. At high $p_T$, particle production is dominated by hard partonic scatterings and fragmentation.

Figures~\ref{cDrawSpectraOOpp} and \ref{cDrawSpectraPbPb} present the $p_T$-differential yields of $\pi^{\pm}$, K$^{\pm}$ and p($\bar{p}$) for pp at $\sqrt{s}$ = 13.6 TeV, and for O–O and Pb–Pb collisions at $\sqrt{s_{NN}}$ = 5.36 TeV, shown for several multiplicity or centrality classes. The upper panels present the $p_T$-differential yields, while the lower panels display the ratios of the spectra in each class to the minimum-bias reference, highlighting relative shape modifications as a function of event multiplicity.
In Pb–Pb collisions, EPOS4 reproduces the characteristic evolution of the spectra with centrality. The spectral slopes become increasingly harder toward central collisions, particularly in the low- and intermediate-$p_{\mathrm{T}}$ regions. 
The ratio-to-minimum-bias show a clear centrality dependence: peripheral collisions exhibit a suppression at intermediate $p_T$, while central events display a clear enhancement, most pronounced for protons. This behavior reflects the strengthening of collective radial expansion with increasing centrality \cite{PhysRevC.88.044910}.
At higher $p_T$, the ratios gradually approach unity within uncertainties, indicating a reduced sensitivity to bulk medium effects and an increasing dominance of hard processes.
O–O collisions exhibit an intermediate behavior between pp and Pb–Pb systems. While signatures of collective dynamics are visible at low and intermediate $p_T$, the spectra remain systematically harder than those in Pb–Pb collisions at comparable scaled multiplicities. 
This trend is clearly reflected in the ratio-to-minimum-bias panels: the enhancement at intermediate $p_T$ in O–O collisions is larger than in Pb–Pb, but smaller than in high-multiplicity pp events. This behaviour highlights the dual nature of O–O collisions, where collective effects coexist with a sizable contribution from non-thermal particle production.
The pp spectra at $\sqrt{s}$ = 13.6 TeV display the hardest slopes among the three systems when evaluated relative to minimum bias, particularly at intermediate and high $p_T$. The ratio-to-minimum-bias distributions show the largest enhancement in high-multiplicity events, reflecting the dominance of hard and semi-hard processes and the comparatively small contribution from thermalized bulk matter.

Additional insight into the observed systematics is provided by the energy dependence of the spectra in Pb–Pb collisions. Figure~\ref{Ratio536by502} shows the ratios of identified-hadron $p_T$ spectra at $\sqrt{s_{NN}}$ = 5.36 TeV to those at 5.02 TeV for three representative centrality classes. For low and intermediate $p_T$, the ratios remain close to unity for all species, indicating weak energy scaling of soft particle production. At high $p_T$, a modest but systematic enhancement is observed at 5.36 TeV, with a clear particle-species dependence: the effect is strongest for pions, followed by kaons, and weakest for protons.

\begin{figure}
	\centering 
	\includegraphics[height=0.44\textheight, angle=0]{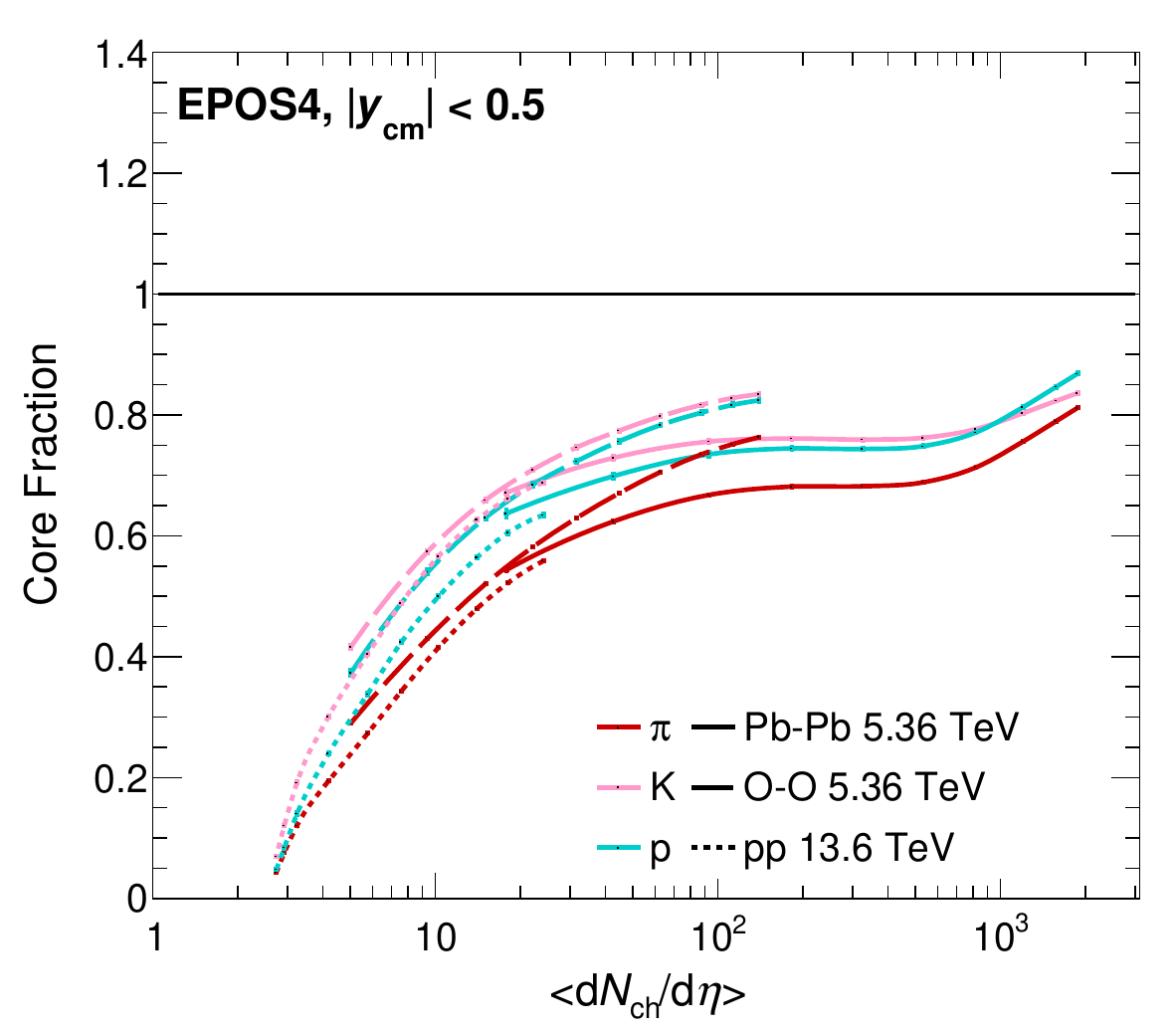}	
	\caption{(Color online) Core fraction of $\pi^{\pm}$, K$^{\pm}$ and p($\bar{p}$) as a function of charged-particle multiplicity for pp collisions at $\sqrt{s}$ = 13.6 TeV, and for O–O and Pb–Pb collisions at $\sqrt{s_{NN}}$ = 5.36 TeV from EPOS4 without the UrQMD hadronic afterburner.} 
	\label{cCoreFraction}%
\end{figure}

A central ingredient of EPOS4 underlying these trends is its dynamical core–corona separation, which regulates the relative contributions of hydrodynamically evolving matter and string-fragmentation–dominated emission. The core fraction, shown in Fig.~\ref{cCoreFraction}, increases monotonically with charged-particle multiplicity for all hadron species and collision systems, reflecting the growing importance of dense, collectively expanding matter at higher event multiplicity. A clear species hierarchy is observed, with $f_{core}(K) > f_{core}(p) > f_{core}(\pi)$. This ordering arises from the different sensitivities of hadron species to core and corona production: pions are abundantly produced via string fragmentation and therefore retain a substantial corona contribution even at high multiplicities, while kaons receive a comparatively enhanced contribution from the thermalized core due to comparatively reduced strange-quark production in the corona component. Protons occupy an intermediate position, benefiting from collective radial flow in the core while still being efficiently produced through baryon–antibaryon pair creation in the corona.
This species-dependent core fraction naturally explains the observed species-dependent hardening of the $p_T$ spectra. As the relative contribution of core-origin hadrons increases with multiplicity and system size, the spectra become progressively harder, particularly for kaons and protons. In pp collisions at $\sqrt{s}$ = 13.6 TeV, the core fraction reaches values of about 60$\%$ in the highest multiplicity classes, indicating a non-negligible contribution from collective dynamics even in small systems. In O–O collisions at $\sqrt{s_{NN}}$ = 5.36 TeV, the core fraction rises rapidly and becomes comparable to that of mid-central Pb–Pb collisions, consistent with the intermediate spectral hardening observed in Fig.~\ref{cDrawSpectraOOpp} and \ref{cDrawSpectraPbPb}. Central Pb–Pb collisions are dominated by core emission, with core fractions exceeding 80$\%$, in line with the strongest collective effects.

\begin{figure}[h]
	\centering 
	\includegraphics[width=1.0\textwidth, angle=0]{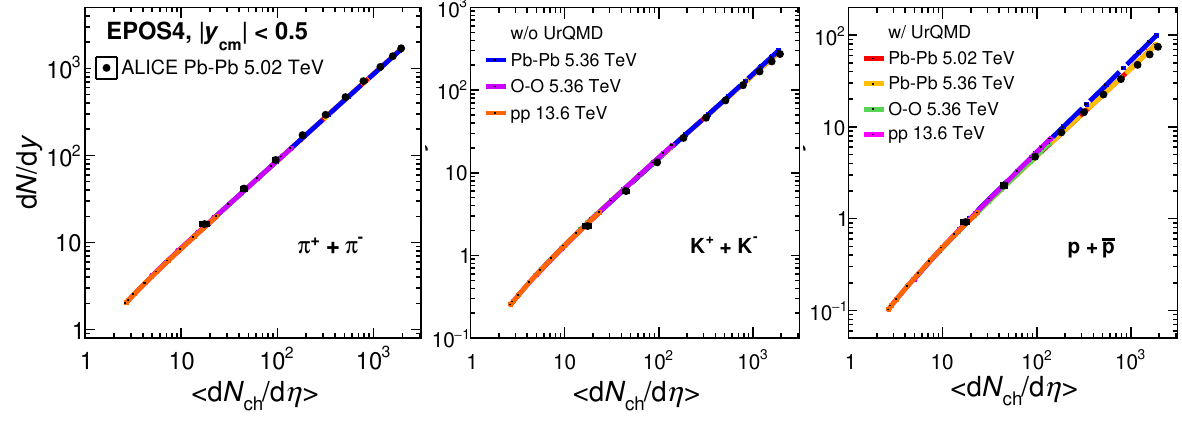}	
	\caption{(Color online) Multiplicity dependence of the midrapidity integrated yields, $dN/dy$, of identified hadrons ($\pi$, $K$, and $p$) as a function of the average charged-particle multiplicity density, $\langle dN_{\mathrm{ch}}/d\eta\rangle$, for pp collisions at $\sqrt{s}=13.6$ TeV and for O–O and Pb–Pb collisions at $\sqrt{s_{NN}}=5.36$ TeV, calculated with EPOS4 for $|y|<0.5$, compared with ALICE data \cite{PhysRevC.101.044907}.} 
	\label{YieldwithNch}%
\end{figure}

\begin{figure}[h]
	\centering 
	\includegraphics[width=1.0\textwidth, angle=0]{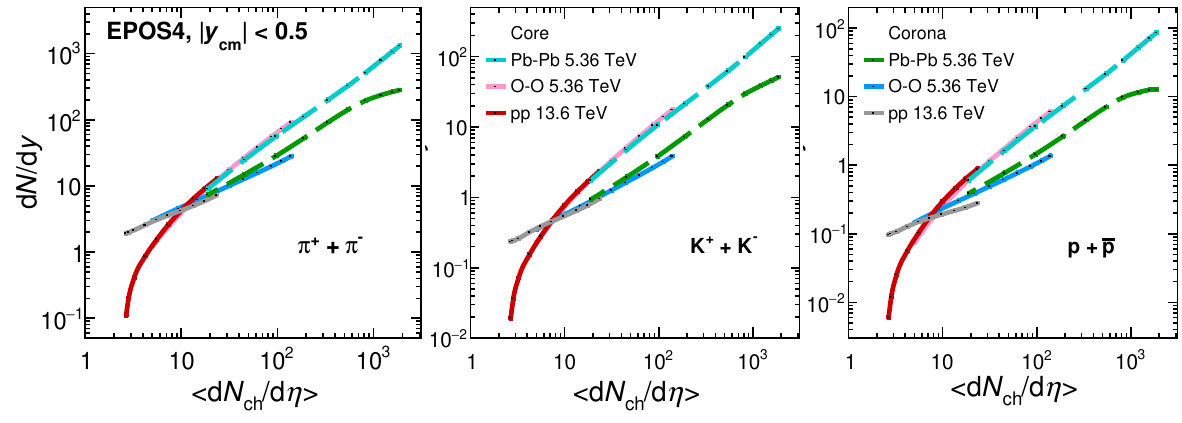}	
	\caption{(Color online) Core and corona contributions to the midrapidity integrated yields, $dN/dy$, of identified hadrons ($\pi$, $K$, and $p$) as a function of $\langle dN_{\mathrm{ch}}/d\eta\rangle$ for pp collisions at $\sqrt{s}=13.6$ TeV and for O–O and Pb–Pb collisions at $\sqrt{s_{NN}}=5.36$ TeV, obtained from EPOS4 calculations for $|y|<0.5$.} 
	\label{YieldwithNchCoreCorona}%
\end{figure}

\subsection{Integrated Yields of Identified Particles}\label{sec6}

The $p_T$-integrated yields of identified hadrons provide essential information on the hadrochemical composition and thermal properties of the system created in high-energy nuclear collisions. 
These yields are sensitive to the chemical freeze-out conditions and reflect the relative contributions of different particle-production mechanisms, including strangeness production and baryon transport \cite{Andronic_2018, 2017}.

Fig. \ref{YieldwithNch} shows the midrapidity integrated yields \(dN/dy\) of \(\pi^{\pm}\), \(K^{\pm}\), and \(p/\overline{p}\) as a function of average charged-particle multiplicity density ($\langle dN_{ch}/d\eta\rangle$) for pp at \(\sqrt{s}\) = 13.6 TeV and for O-O and Pb–Pb collisions at \(\sqrt{s_{NN}}\) = 5.36 TeV, obtained from EPOS4 calculations with and without hadronic re-scattering. 
For comparison, available ALICE measurements for Pb-Pb collisions at 5.02 TeV are overlaid \cite{PhysRevC.101.044907}.
At Run 2 energies, the EPOS4 provides a good description of the experimental data, reproducing both the absolute yields and the observed hierarchy \( \pi > K > p \) across a wide multiplicity range.
At \(\sqrt{s_{NN}} = 5.36\) TeV, the model predicts only a negligible increase in the integrated yields for all particle species relative to \(\sqrt{s_{NN}} = 5.02\) TeV.
The inclusion of hadronic re-scattering has a negligible impact on pion and kaon yields, while a small suppression of the proton yield is observed towards central Pb-Pb collisions, consistent with baryon-antibaryon annihilation effects in the hadronic phase \cite{PhysRevC.89.044911}.

Further insight into the origin of the integrated yields is obtained by separating the contributions from core and corona particle production, as shown in Fig.~\ref{YieldwithNchCoreCorona}. Both contributions exhibit a non-linear increase with $\langle dN_{ch}/d\eta\rangle$, with the relative contribution from the core rising steadily toward higher multiplicities. In low-multiplicity pp and O–O collisions, particle production is dominated by the corona component, reflecting string-fragmentation–driven hadronization. With increasing multiplicity, core emission becomes increasingly important, particularly in O–O and Pb–Pb collisions.
The transition from corona-dominated to core-dominated production occurs at different multiplicity scales for different particle species. For kaons, the crossover takes place at lower multiplicities than for pions, consistent with the enhanced production of strange hadrons in the thermalized core. In contrast, protons exhibit a crossover at higher multiplicities than kaons, reflecting the absence of strangeness content despite their larger mass. This behaviour indicates that both particle mass and strangeness play a crucial role in determining the multiplicity scale at which core hadronization becomes dominant.

\begin{figure}
	\centering 
	\includegraphics[width=\textwidth, angle=0]{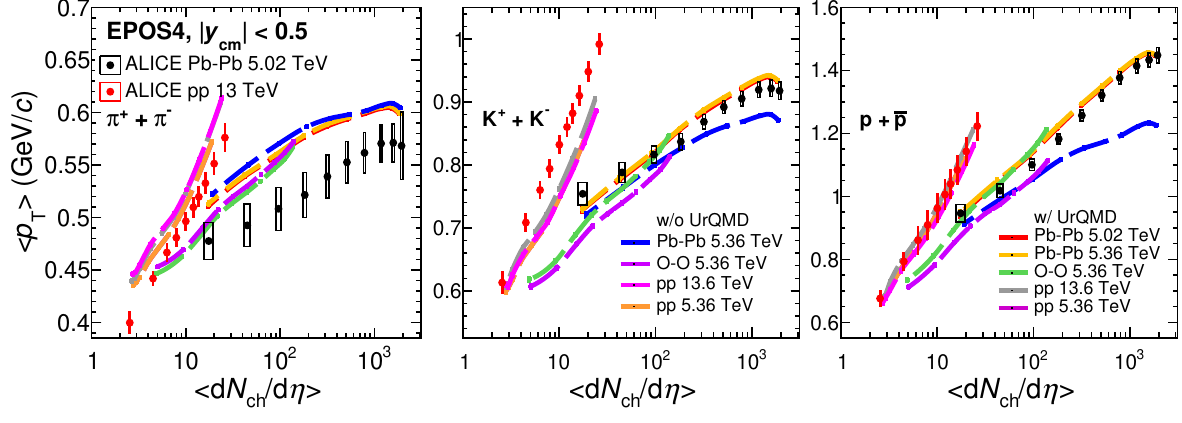}	
	\caption{(Color online) Multiplicity dependence of the mean transverse momentum, $\langle p_T \rangle$, of identified hadrons ($\pi$, $K$, and $p$) at midrapidity ($|y_{cm}|<0.5$) in pp collisions at $\sqrt{s}=13.6$ TeV and in O–O and Pb–Pb collisions at $\sqrt{s_{NN}}=5.36$ TeV. Results for Pb–Pb collisions at $\sqrt{s_{NN}}=5.02$ TeV from EPOS4 and available ALICE measurements are also shown for comparison \cite{Acharya_2020, PhysRevC.101.044907}. EPOS4 calculations are presented with and without UrQMD hadronic re-scattering.} 
	\label{meanptwithNch}%
\end{figure}

\begin{figure}
	\centering 
	\includegraphics[width=\textwidth, angle=0]{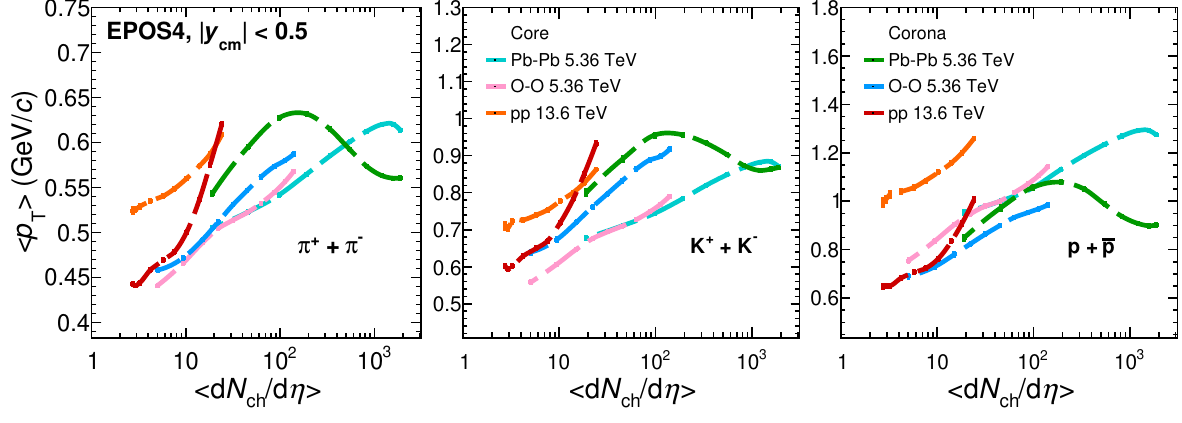}	
	\caption{(Color online) Multiplicity dependence of the mean transverse momentum, $\langle p_T \rangle$, of identified hadrons ($\pi$, $K$, and $p$) at midrapidity ($|y_{cm}|<0.5$) in pp collisions at $\sqrt{s}=13.6$ TeV and in Pb–Pb and O–O collisions at $\sqrt{s_{NN}}=5.36$ TeV, showing the separate contributions from the hydrodynamic core and the string-fragmentation-dominated corona components in EPOS4.} 
	\label{meanpt536withNch}%
\end{figure}

\subsection{Mean Transverse Momentum of Identified Particles}\label{sec7}

The mean transverse momentum, $\langle p_T \rangle$, of identified particles is a key observable for disentangling the interplay between thermal motion, collective expansion and non-thermal particle production mechanisms in high-energy nuclear collisions.
It encodes information on both the effective temperature of the system and the strength of radial flow developed during its evolution.
A well-established feature in heavy-ion collisions is the mass ordering of $\langle p_T \rangle$, whereby heavier hadrons acquire larger mean transverse momenta, reflecting the stronger response to collective expansion.
This behaviour has been firmly observed in Pb–Pb collisions at the LHC, providing strong evidence for hydrodynamic expansion of the medium \cite{PhysRevC.88.044910}.
In contrast, for small collision systems such as pp and the recently accessible O–O collisions, the presence of genuine collective flow remains under debate \cite{Nagle_2018}.
While QCD-inspired event generators such as AMPT \cite{PhysRevC.72.064901} and PYTHIA \cite{Sj_strand_2015} can reproduce radial-flow–like patterns through final-state mechanisms (e.g. color reconnection in PYTHIA), these approaches do not rely on hydrodynamic evolution \cite{Christiansen_2015}.
In this context, QGP-inspired EPOS4 model, which combines microscopic string dynamics with macroscopic hydrodynamics, provides a well-suited framework to investigate the system-size dependence of $\langle p_T \rangle$ across pp, O–O, and Pb–Pb collisions \cite{PhysRevC.109.014910}.

The $\langle p_T \rangle$ of charged pions ($\pi^{\pm}$), kaons ($K^{\pm}$) and (anti)protons ($p, \overline{p}$) at midrapidity ($|y| < 0.5$) is studied as a function of average charged-particle multiplicity density ($\langle dN_{ch}/d\eta \rangle$) for pp collisions at $\sqrt{s}$ = 13.6 TeV and for O-O and Pb–Pb collisions at $\sqrt{s_{NN}}$ = 5.36 TeV using EPOS4 with and without UrQMD hadronic re-scattering. 
For comparison, EPOS4 results for Pb-Pb collisions at $\sqrt{s_{NN}}$ = 5.02 TeV and available ALICE measurements are also shown in Fig.~\ref{meanptwithNch} \cite{Acharya_2020, PhysRevC.101.044907}.
For each collision system and particle species, $\langle p_T \rangle$ shows an overall increase with $\langle dN_{ch}/d\eta \rangle$, indicating a growing contribution from harder particle production mechanisms with increasing event multiplicity. At a fixed multiplicity, however, clear system-dependent differences are observed: pp collisions exhibit systematically larger $\langle p_T \rangle$ values than O–O and Pb–Pb collisions. This demonstrates that $\langle p_T \rangle$ does not follow a universal scaling with multiplicity across systems and reflects the different microscopic origins of high multiplicity. In small systems, such as pp, high multiplicity is often driven by rare hard and semi-hard processes, whereas in heavy-ion collisions it predominantly arises from multiple soft interactions within an extended medium. O–O collisions exhibit an intermediate behaviour between pp and Pb–Pb, reinforcing their role as a bridge system connecting small and large collision systems. A clear mass dependence is observed across all systems, with $\langle p_T \rangle(\pi) < \langle p_T \rangle(K) < \langle p_T \rangle(p)$ over the full multiplicity range. This hierarchy is consistent with the combined effects of collective radial expansion and the increasing contribution of heavier hadrons at higher transverse momenta.  
The inclusion of hadronic re-scattering via UrQMD leads to a moderate increase in $\langle p_{\rm{T}} \rangle$ for most species, with the effect being most pronounced for protons due to baryon–antibaryon annihilation in the hadronic phase. In contrast, pion $\langle p_{\rm{T}} \rangle$ remains largely unaffected.
Overall, EPOS4 provides a good description of the ALICE data for Pb-Pb collisions at $\sqrt{s_{NN}}$ = 5.02 TeV, although it slightly overestimates $\langle p_{\rm{T}} \rangle$ for pions.

Further insight is obtained by separating the hydrodynamically evolving core and string-fragmentation-dominated corona contributions, as shown in Fig.~\ref{meanpt536withNch}. 
In pp and O–O collisions, the corona contribution to $\langle p_T \rangle$ dominates over the core across the full multiplicity range, highlighting the importance of non-thermal particle production in small systems. In Pb-Pb collisions, the corona dominates at low and intermediate multiplicities, while the core contribution becomes increasingly significant toward high multiplicity events. For protons, the core contribution is already significant at relatively low multiplicities, reflecting their stronger coupling to the collectively expanding medium.
At comparable $\langle dN_{ch}/d\eta \rangle$, the corona $\langle p_T \rangle$ follows a clear hierarchy, with the largest values observed in pp collisions, followed by Pb–Pb and then O–O collisions. In contrast, the core $\langle p_T \rangle$ in O–O and Pb–Pb collisions evolves smoothly and similarly with multiplicity, whereas pp collisions show systematically higher core $\langle p_T \rangle$ values. This demonstrates that identical final-state multiplicities can originate from qualitatively different microscopic mechanisms depending on the system size, and that $\langle p_T \rangle$ retains sensitivity to the underlying collision dynamics beyond a simple multiplicity scaling.

\begin{figure}[h]
	\centering 
	\includegraphics[width=0.8\textwidth, angle=0]{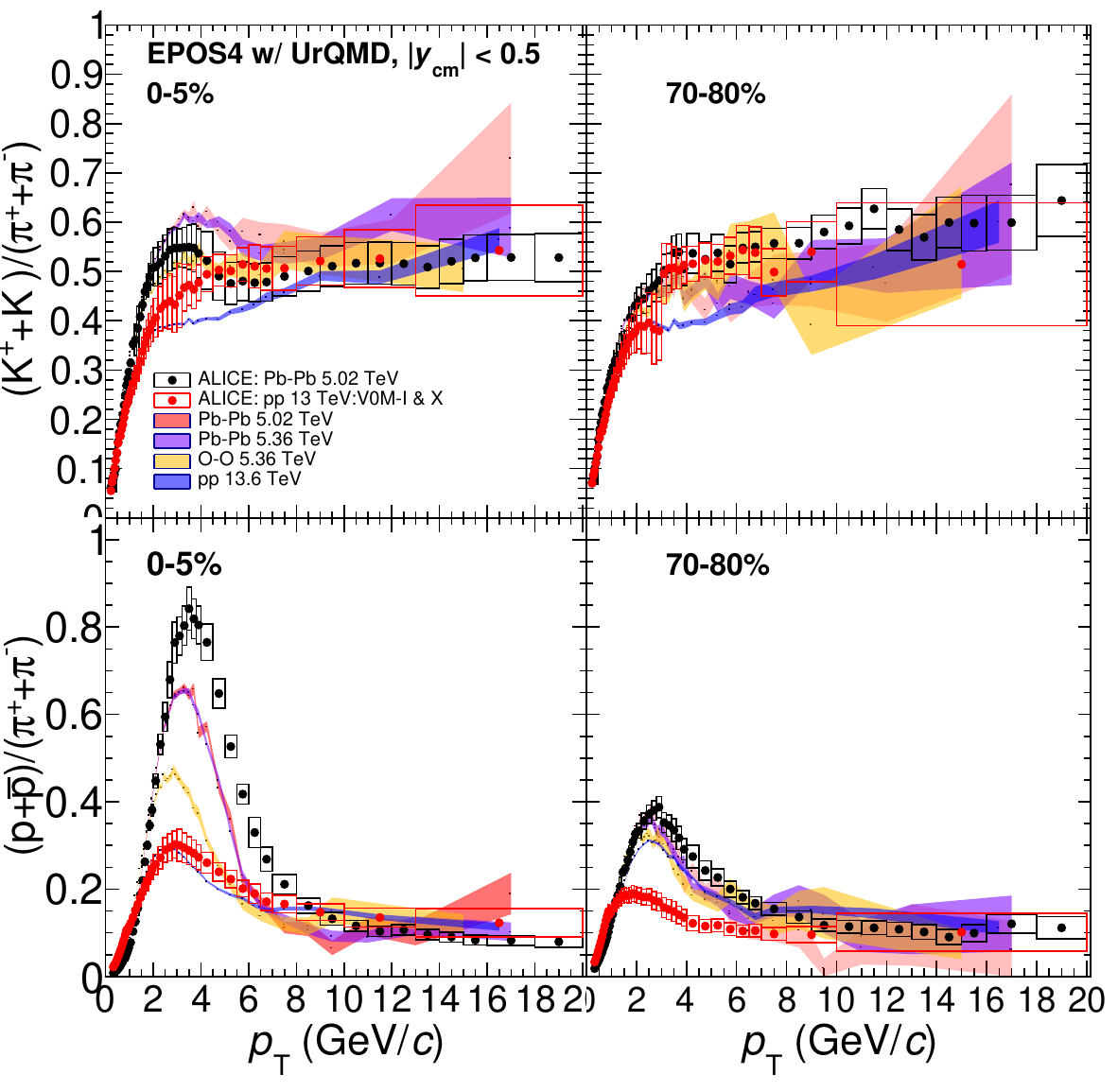}	
	\caption{(Color online) Transverse-momentum–differential particle ratios $(K^+ + K^-)/(\pi^+ + \pi^-)$ (top) and $(p + \bar{p})/(\pi^+ + \pi^-)$ (bottom) for central (0–5$\%$) and peripheral (70-80$\%$) event classes in pp collisions at $\sqrt{s}=13.6$ TeV, and in O–O and Pb–Pb collisions at $\sqrt{s_{NN}}=5.36$ TeV. Results for Pb–Pb collisions at $\sqrt{s_{NN}}=5.02$ TeV and pp collisions at $\sqrt{s}=13$ TeV from ALICE are shown for comparison \cite{Acharya_2020, PhysRevC.101.044907}. EPOS4 calculations include the UrQMD hadronic afterburner.} 
	\label{ratiowPt}%
\end{figure}

\begin{figure}[h]
	\centering 
	\includegraphics[width=1.0\textwidth, angle=0]{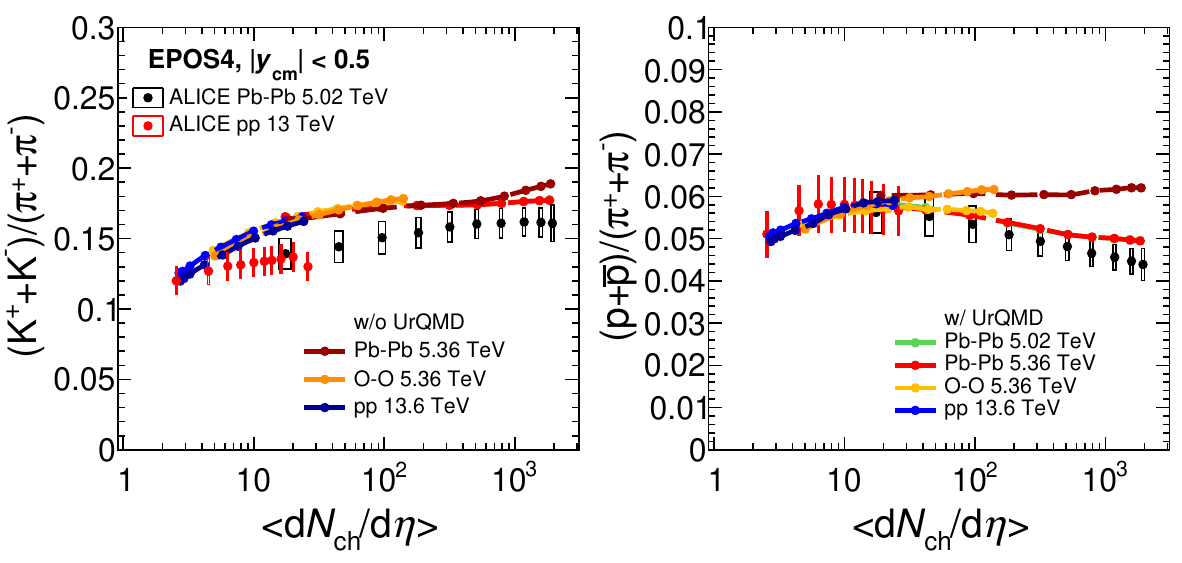}	
    \includegraphics[width=1.0\textwidth, angle=0]{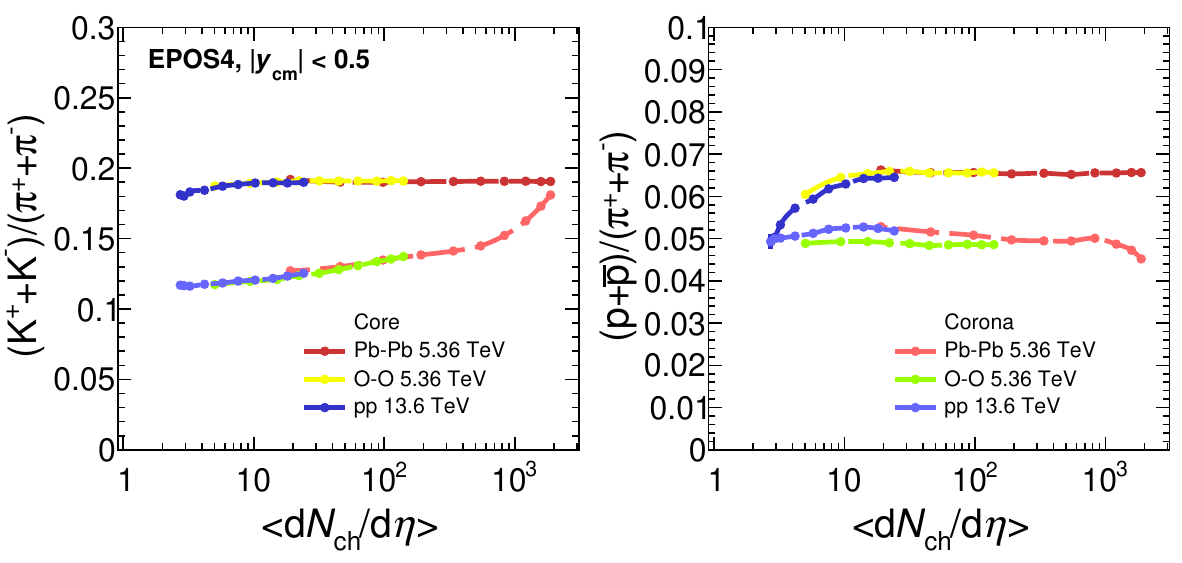}	
	\caption{(Color online) Multiplicity dependence of $p_T$-integrated particle ratios $K/\pi$ (left) and $p/\pi$ (right) as functions of $\langle dN_{ch}/d\eta \rangle$ in pp collisions at $\sqrt{s}=13.6$ TeV and in O–O and Pb–Pb collisions at $\sqrt{s_{NN}}=5.36$ TeV. Available ALICE measurements for pp collisions at $\sqrt{s}=13$ TeV and Pb–Pb collisions at $\sqrt{s_{NN}}=5.02$ TeV are shown for comparison \cite{Acharya_2020, PhysRevC.101.044907}. The lower panels show the corresponding core and corona contributions in EPOS4.} 
	\label{RatioWMult}%
\end{figure}

\begin{figure}[h]
	\centering 
	\includegraphics[width=0.8\textwidth, angle=0]{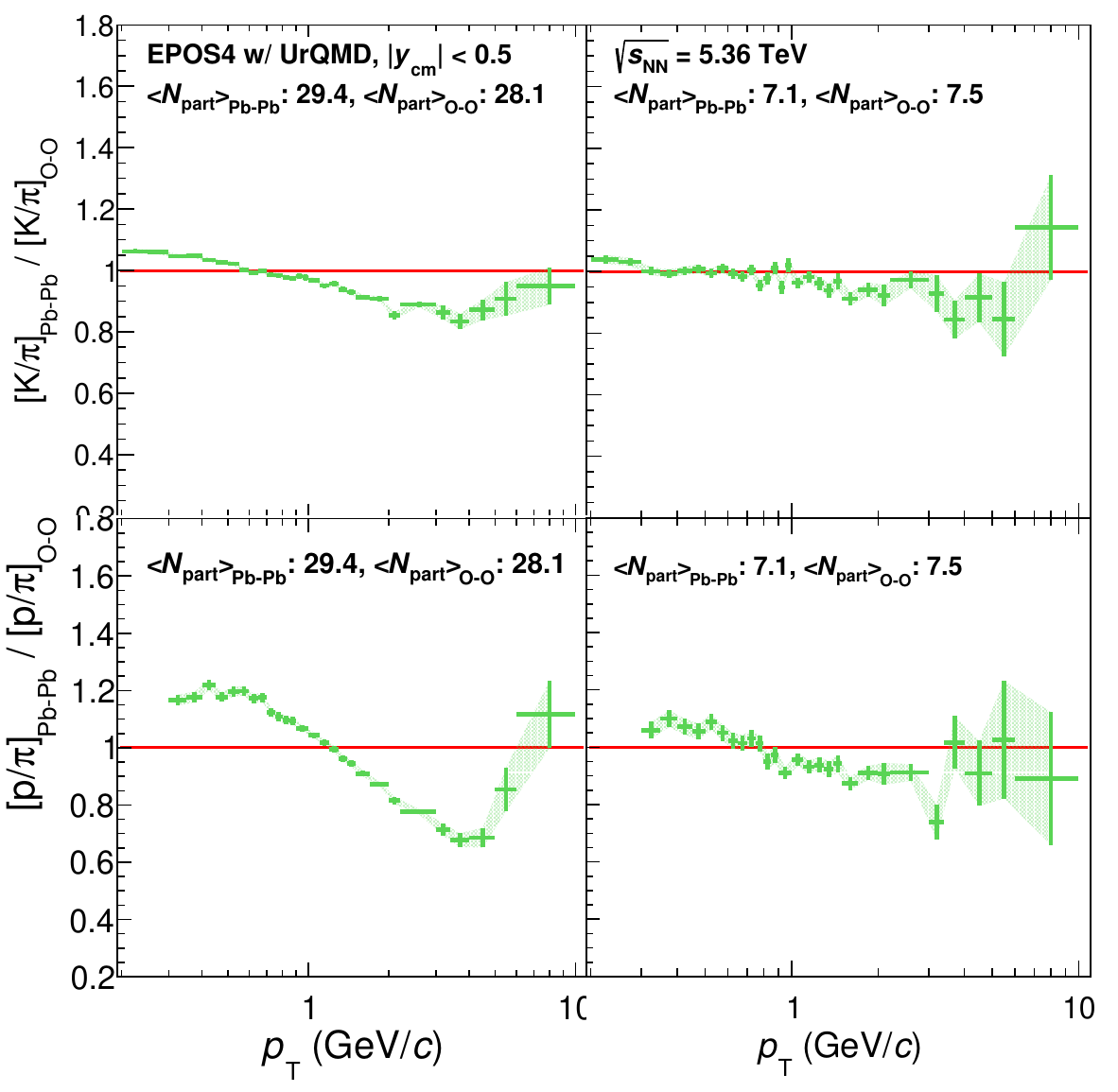}	
	\caption{(Color online) Double ratios of $K/\pi$ (top) and $p/\pi$ (bottom) in Pb–Pb to O-O collisions at comparable average numbers of participating nucleons, $\langle N_{part} \rangle$, as functions of transverse momentum at $\sqrt{s_{NN}}=5.36$ TeV. EPOS4 calculations include the UrQMD hadronic afterburner.} 
	\label{doubleratiowPt}%
\end{figure}

\subsection{Particle Ratios and Core–Corona Contributions}\label{sec8}

Ratios of identified-hadron production provide sensitive probes of the underlying particle production mechanisms in high-energy nuclear collisions. In particular, the meson-to-meson (K$/\pi$) and baryon-to-meson (p$/\pi$) ratios reflect the interplay between collective expansion, hadronization dynamics, and fragmentation processes, and therefore offer complementary insight into the hadrochemical evolution of the system \cite{PhysRevC.101.044907, Acharya_2020}. 
Studying these ratios as functions of transverse momentum and charged-particle multiplicity enables a differential characterization of relative contributions from the hydrodynamically evolving core and the fragmentation-dominated corona.
Baryon-to-meson ratios such as p$/\pi$ are especially informative, as they are sensitive not only to particle mass but also to the underlying hadronization mechanism. Enhancements of $p/\pi$ at intermediate transverse momentum are commonly associated with collective radial flow and/or recombination effects \cite{Fries_2008}, whereas meson-to-meson ratios such as $K/\pi$ are expected to be less sensitive to collective dynamics and more strongly influenced by flavor production in fragmentation \cite{2017}.

Figure \ref{ratiowPt} presents the \(p_{\rm{T}}\)-differential ratios $K/\pi=(K^++K^-)/(\pi^++\pi^-)$ and $p/\pi=(p+\bar{p})/(\pi^++\pi^-)$ for 0-5$\%$ (most central) and 70-80$\%$ (peripheral) event classes in pp collisions at $\sqrt{s}=$ 13.6 TeV, and in O-O and Pb$-$Pb collisions at $\sqrt{s_{\rm NN}}=$ 5.36 TeV. 
EPOS4 calculations including the UrQMD hadronic afterburner are compared to available ALICE measurements for pp collisions at 13 TeV and Pb-Pb collisions at 5.02 TeV \cite{Acharya_2020, PhysRevC.101.044907}. 
The p/$\pi$ ratio exhibits a pronounced enhancement at intermediate transverse momenta ($p_T \sim 2$--$4$~GeV/$c$), whose magnitude and peak position increase systematically with event multiplicity and system size. The enhancement is weakest in pp collisions, becomes more pronounced in O–O collisions, and reaches its maximum in central Pb–Pb collisions, reflecting the increasing importance of collective radial expansion from small to large systems. At high $p_{\rm{T}}$, the p/$\pi$ ratios for different systems and multiplicity classes gradually converge within uncertainties, indicating a reduced sensitivity to bulk medium effects and the dominance of fragmentation-driven particle production. While EPOS4 reproduces the overall shape and magnitude of the K/$\pi$ ratio reasonably well, it underestimates the p/$\pi$ ratio in the intermediate-$p_{\rm{T}}$ region, particularly in central Pb-Pb collisions, suggesting limitations in the model description of baryon production in this momentum range.
In contrast, the $K/\pi$ ratio shows a comparatively weak dependence on $p_T$ above $\sim$3-4 GeV/$c$ and only a mild sensitivity to event multiplicity across all systems.
This behaviour indicates that strange-meson production at intermediate and high $p_{\rm{T}}$ is largely governed by fragmentation processes, where differences in hadron mass play a subdominant role compared to quark-flavour composition. The contrasting behaviours of $p/\pi$ and $K/\pi$ thus highlight the stronger sensitivity of baryon production to collective dynamics.

The upper panels of Fig. \ref{RatioWMult} presents the $p_{\rm{T}}$-integrated K/$\pi$ and p/$\pi$ ratios as a function of the average charged-particle multiplicity density, $\langle dN_{ch}/d\eta \rangle$, for pp collisions at $\sqrt{s}=$ 13.6 TeV, and for O-O and Pb$-$Pb collisions at $\sqrt{s_{\rm NN}}=$ 5.36 TeV, together with available ALICE measurements \cite{Acharya_2020, PhysRevC.101.044907}. 
Clear differences emerge between EPOS4 calculations with and without the UrQMD hadronic afterburner.
For the p/$\pi$ ratio, EPOS4 with UrQMD closely follows the Run 2 ALICE trend: the ratio increases from low-multiplicity pp events, evolves smoothly through the O–O collisions, and decreases toward high-multiplicity Pb–Pb collisions. This suppression at large multiplicities reflects the increasing importance of hadronic-phase effects, in particular baryon–antibaryon annihilation, in dense and long-lived systems.
In contrast, calculations without UrQMD show a smoother evolution of p/$\pi$ with multiplicity and do not exhibit a significant decrease at high Pb–Pb multiplicities, indicating that the observed suppression is primarily driven by hadronic re-scattering. In pp collisions, the difference between calculations with and without UrQMD is minimal, consistent with the short lifetime of the hadronic phase in small systems.
A similar but weaker system-size dependence is observed for the K/$\pi$ ratio. In EPOS4, K/$\pi$ increases from low-multiplicity pp events and evolves smoothly through O–O to Pb–Pb collisions, approaching a nearly constant value at high multiplicities. Deviations between calculations with and without UrQMD are small over most of the multiplicity range and become noticeable only in the most central Pb–Pb collisions.
Modest discrepancies with ALICE data at low-multiplicity Pb–Pb and high-multiplicity pp events indicate the limitations of a purely multiplicity-driven description and underline the role of system size and collision geometry.

Additional insight is obtained by separating the multiplicity-integrated particle ratios into contributions from the hydrodynamically evolving core and the string-fragmentation dominated corona, as shown in the lower panels of Fig.~\ref{RatioWMult}. 
For the core component, the p/$\pi$ ratio increases from pp to low-multiplicity O–O collisions and then saturates, remaining approximately constant up to high-multiplicity Pb–Pb events, indicating that baryon production in the thermalized core becomes efficient already at moderate event multiplicity.
For the corona component, the p/$\pi$ ratio exhibits only weak variations with multiplicity, consistent with baryon production dominated by fragmentation.
The core contribution to the K/$\pi$ ratio is nearly flat across all systems, while the corona contribution increases smoothly from pp to O–O and Pb–Pb collisions, reflecting enhanced strange-quark production in fragmentation with increasing effective string mass.

Figure~\ref{doubleratiowPt} compares the K/$\pi$ and p/$\pi$ ratios in O–O and Pb–Pb collisions at similar average numbers of participating nucleons, $\langle N_{part} \rangle$, at $\sqrt{s_{NN}}=5.36$~TeV. At low $p_T$, the double ratios exceed unity, while in the intermediate-$p_T$ region they fall below unity, before becoming compatible with unity again at high $p_T$ within uncertainties. This behavior indicates that, once geometrical effects are accounted for through participant scaling, the relative hadrochemical composition in O–O and Pb–Pb collisions at comparable $\langle N_{part} \rangle$ are largely governed by similar underlying dynamics, with residual differences arising from system-size–dependent corona contributions and enhanced fluctuations in the smaller O–O system.

\begin{figure}
	\centering 
	\includegraphics[width=0.5\textwidth, angle=0]{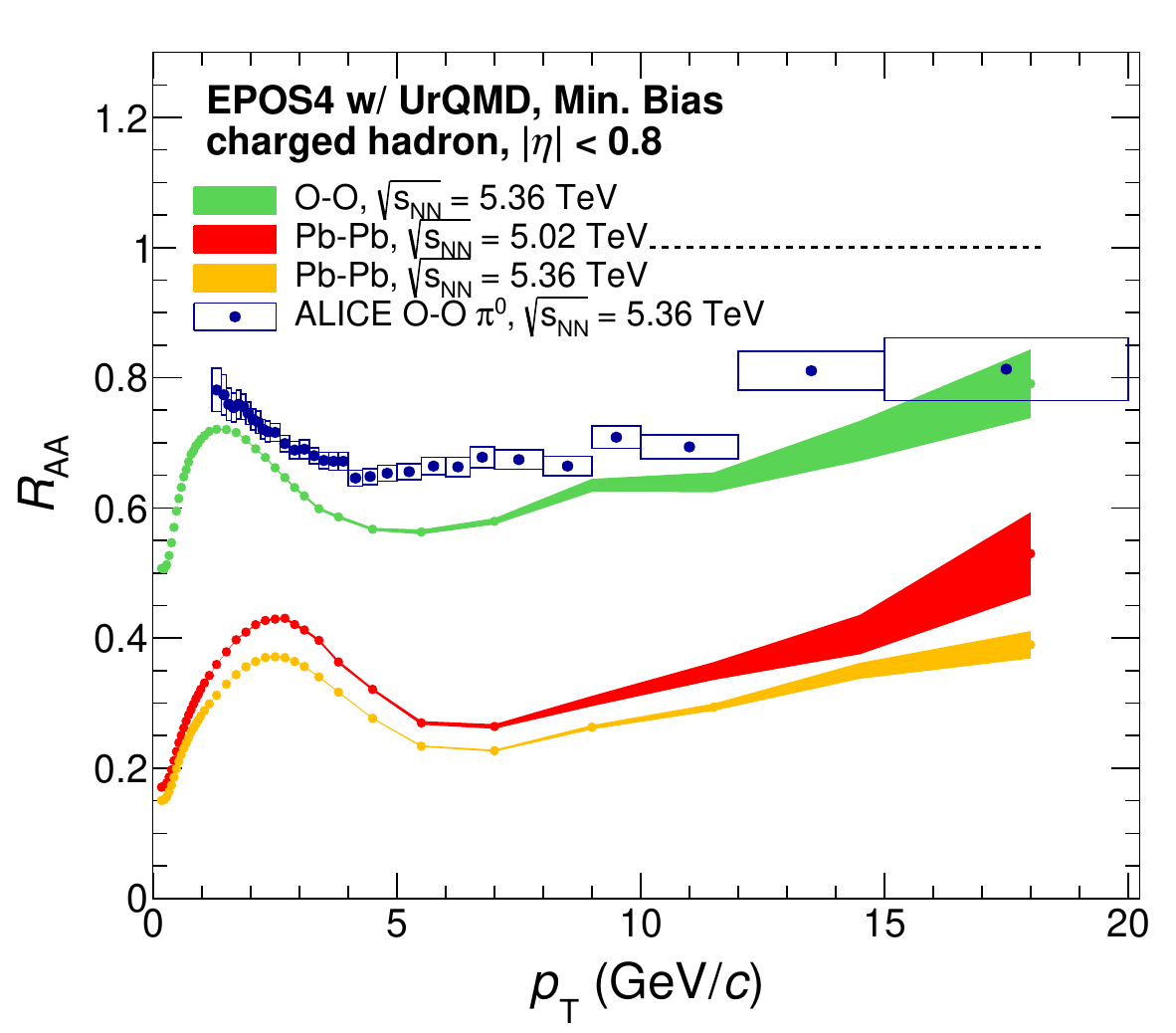}	
	\caption{(Color online) Nuclear modification factor $R_{AA}$ of charged hadrons as a function of transverse momentum in O–O and Pb–Pb collisions from EPOS4 with UrQMD contributions. ALICE preliminary $\pi^0$ measurement is shown for reference \cite{RAA_OO}.} 
	\label{cRAAcharged}%
\end{figure}

\begin{figure}
	\centering 
	\includegraphics[width=0.8\textwidth, angle=0]{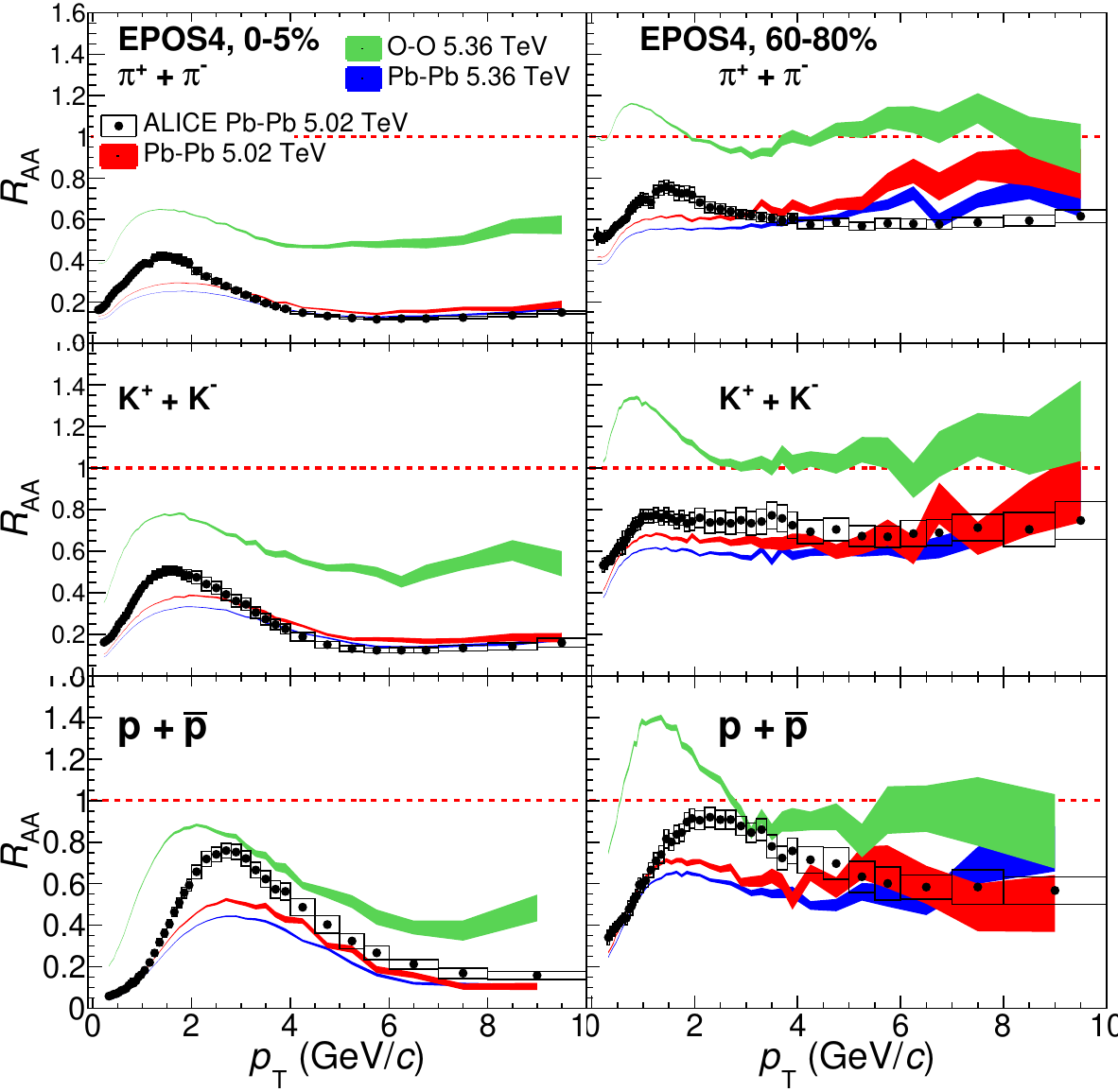}	
	\caption{(Color online) Transverse-momentum–differential nuclear modification factor $R_{AA}$ of identified hadrons ($\pi^{\pm}$, $K^{\pm}$, and $p+\bar{p}$) in central
(0–5\%) and peripheral (60–80\%) O–O and Pb–Pb collisions at
$\sqrt{s_{NN}} = 5.36$ TeV from EPOS4 with the UrQMD hadronic afterburner.
Results for Pb–Pb collisions at $\sqrt{s_{NN}} = 5.02$ TeV and corresponding
ALICE measurements are shown for comparison \cite{PhysRevC.101.044907}.} 
	\label{RAA}%
\end{figure}

\begin{figure}
	\centering 
	\includegraphics[width=0.5\textwidth, angle=0]{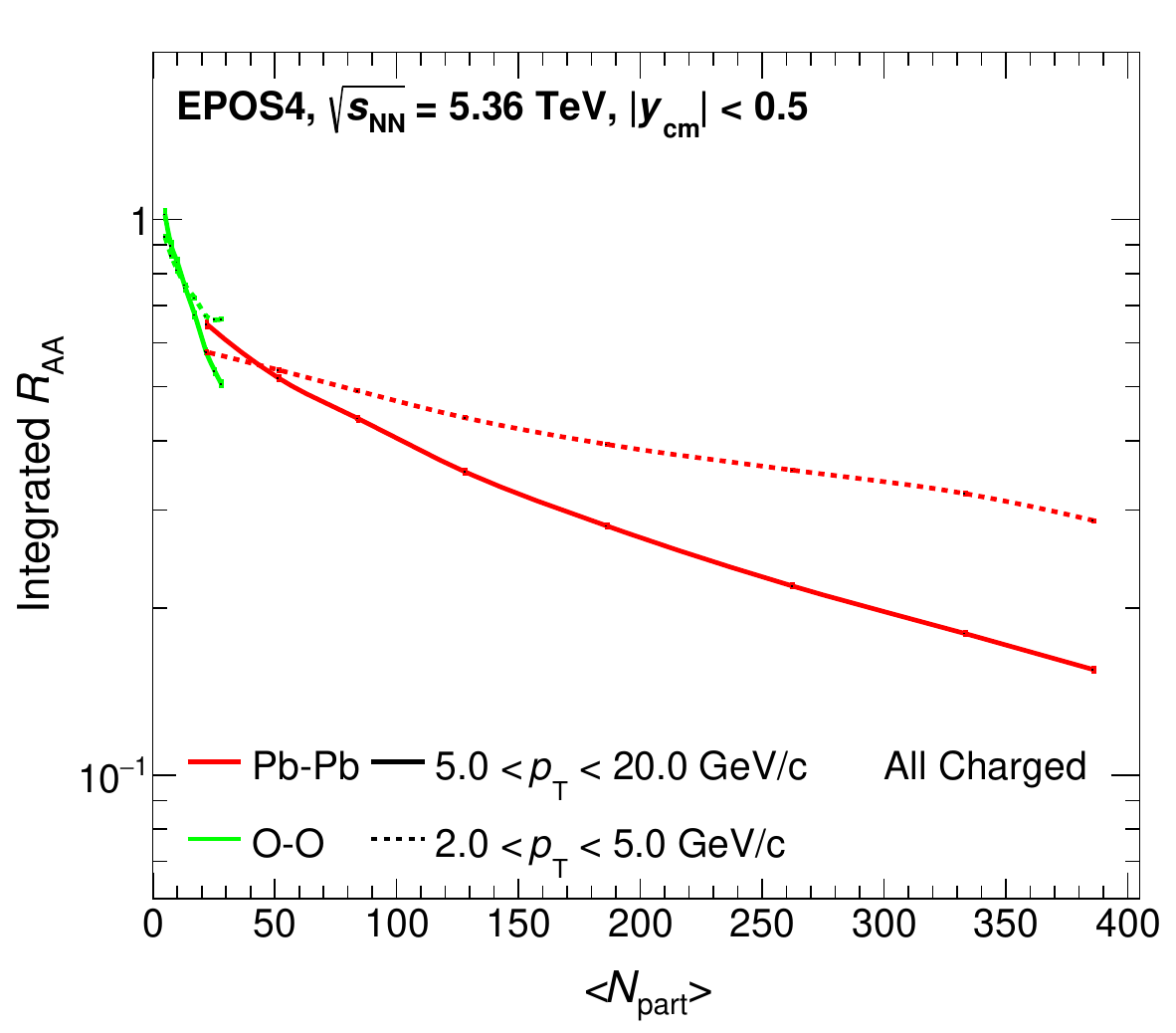}	
	\caption{(Color online) $p_T$-integrated nuclear modification factor of charged hadrons
as a function of the average number of participating nucleons,
$\langle N_{\text{part}} \rangle$, in O–O and Pb–Pb collisions at
$\sqrt{s_{NN}} = 5.36$ TeV. Results are shown separately for the intermediate-$p_T$
($2 < p_T < 5$ GeV/$c$) and high-$p_T$ ($5 < p_T < 20$ GeV/$c$) intervals.} 
	\label{cRAAIntegrated}%
\end{figure}

\subsection{Nuclear Modification Factor of Identified Hadrons}\label{subsec7}

The nuclear modification factor, $R_{AA}$, provides a quantitative measure of medium-induced modifications of particle production in the nucleus–nucleus (A–A) collisions relative to an incoherent superposition of proton-proton (pp) interactions. 
It is defined as

$$
R_{AA}(p_T) = \frac{1}{\langle N_{\text{coll}} \rangle} \cdot \frac{d^2N_{AA}/dp_Tdy}{d^2N_{pp}/dp_Tdy},
$$

where $\langle N_{\text{coll}} \rangle$ denotes the average number of binary nucleon-nucleon collisions for a given centrality class.
In the absence of nuclear effects, particle production is expected to scale with $\langle N_{\text{coll}} \rangle$, yielding $R_{AA}$ = 1.
Deviations from unity therefore provide direct evidence of medium effects. In particular, suppression at intermediate and high transverse momentum is commonly interpreted as partonic energy loss in the quark–gluon plasma (QGP), while modifications at low $p_T$ reflect collective flow and hadronization dynamics \cite{PhysRevC.101.044907}.

In this study, $R_{AA}$ is evaluated using EPOS4 calculations including the UrQMD hadronic afterburner. The pp reference spectra are consistently taken from EPOS4 at the corresponding collision energies.
Figure \ref{cRAAcharged} presents the transverse-momentum dependence of the nuclear modification factor for inclusive charged hadrons. A pronounced suppression is observed in both collision systems over a broad $p_T$ range. The suppression is significantly stronger in Pb–Pb than in O-O collisions and increases toward high $p_T$, consistent with path-length-dependent partonic energy loss. The EPOS4 with UrQMD predictions for O–O are compared with available ALICE preliminary measurements and are found to be compatible within uncertainties. The similarity of the suppression pattern to that observed for neutral pions is consistent with expectations from partonic energy loss and a weak hadron-species dependence within current precision.

Figure \ref{RAA} shows the $p_T$-differential $R_{AA}$ for identified hadrons ($\pi^{\pm}$, $K^{\pm}$, $p$, $\overline{p}$) for central (0–5$\%$) and peripheral (60-80$\%$) O-O and Pb–Pb collisions at $\sqrt{s_{NN}} = 5.36$ TeV. For reference, EPOS4 calculations and available ALICE measurements for Pb-Pb collisions at $\sqrt{s_{NN}} =$5.02 TeV are also shown \cite{PhysRevC.101.044907}.
For Pb-Pb collisions at $\sqrt{s_{NN}} = 5.02$ TeV, EPOS4 reproduces the main qualitative features observed by ALICE: a suppression of particle production over the full measured $p_T$ range, reaching a minimum around $p_T \sim 6$–8 GeV/$c$, followed by a gradual rise toward higher $p_T$. 
Quantitatively, however, EPOS4 tends to predict a somewhat stronger suppression
than observed in the data in the intermediate-$p_T$ region for all particle species.
At low-$p_T$, a clear particle-species hierarchy, $R_{AA}^{(p)} > R_{AA}^{(K)} > R_{AA}^{(\pi)}$, is observed, reflecting the combined effects of hadron mass, spectral shapes, and collective radial expansion.
At $\sqrt{s_{NN}}=5.36$ TeV, EPOS4 predicts a slightly stronger suppression at intermediate and high $p_T$ for all particle species compared to 5.02 TeV, consistent with enhanced partonic energy loss in a hotter and denser medium. The particle-species hierarchy persists, indicating that mass-dependent collective effects remain relevant also at the higher collision energy.
A clear centrality dependence is observed in Pb–Pb collisions, with central events exhibiting systematically stronger suppression than peripheral ones over the full $p_T$ ranges, consistent with the increased path length and medium density in more central collisions.
The inclusion of O–O collisions at $\sqrt{s_{NN}}=5.36$ TeV provides access to medium effects in an intermediate system size \cite{cmscollaboration2025discoverysuppressedchargedparticleproduction}. In central O–O collisions, a suppression of $R_{AA}$ is observed at intermediate and high $p_T$ for all particle species, although its magnitude is systematically weaker than in Pb–Pb collisions, reflecting the reduced system size and shorter in-medium path lengths. At low $p_T$, the proton-to-pion hierarchy remains visible, indicating the presence of collective effects even in this lighter collision system. Peripheral O–O collisions exhibit a substantially reduced suppression, consistent with diminishing medium effects.

The system-size dependence of medium-induced suppression is further illustrated in Fig. \ref{cRAAIntegrated}, which presents the $p_T$-integrated nuclear modification factor of charged hadrons as a function of the average number of participating nucleons, $\langle N_{\text{part}} \rangle$, in O–O and Pb–Pb collisions at $\sqrt{s_{NN}}=5.36$ TeV. 
Results are presented separately for the intermediate-$p_T$ ($2 < p_T < 5$~GeV/$c$) and high-$p_T$ ($5 < p_T < 20$~GeV/$c$) regions.
In O--O collisions, a suppression is already observed in both
$p_T$ intervals and decreases gradually with increasing $\langle N_{\text{part}} \rangle$, indicating the progressive development of medium effects even in this light system.
In Pb--Pb collisions, the suppression becomes significantly stronger and decreases monotonically toward central events, reaching substantially lower values than in O--O collisions over a much broader $\langle N_{\text{part}} \rangle$ range. The stronger reduction observed at high $p_T$ in both systems reflects the enhanced sensitivity of hard probes to the increasing medium density and path length, in agreement with expectations from radiative and collisional partonic energy-loss mechanisms.
No indication of saturation of the integrated $R_{AA}$ is observed within the explored $\langle N_{\text{part}} \rangle$ range, suggesting that medium-induced suppression continues to strengthen with increasing system size and collision centrality \cite{PhysRevLett.101.232301}.

\begin{figure}[h]
    \centering 

    \includegraphics[width=\textwidth]{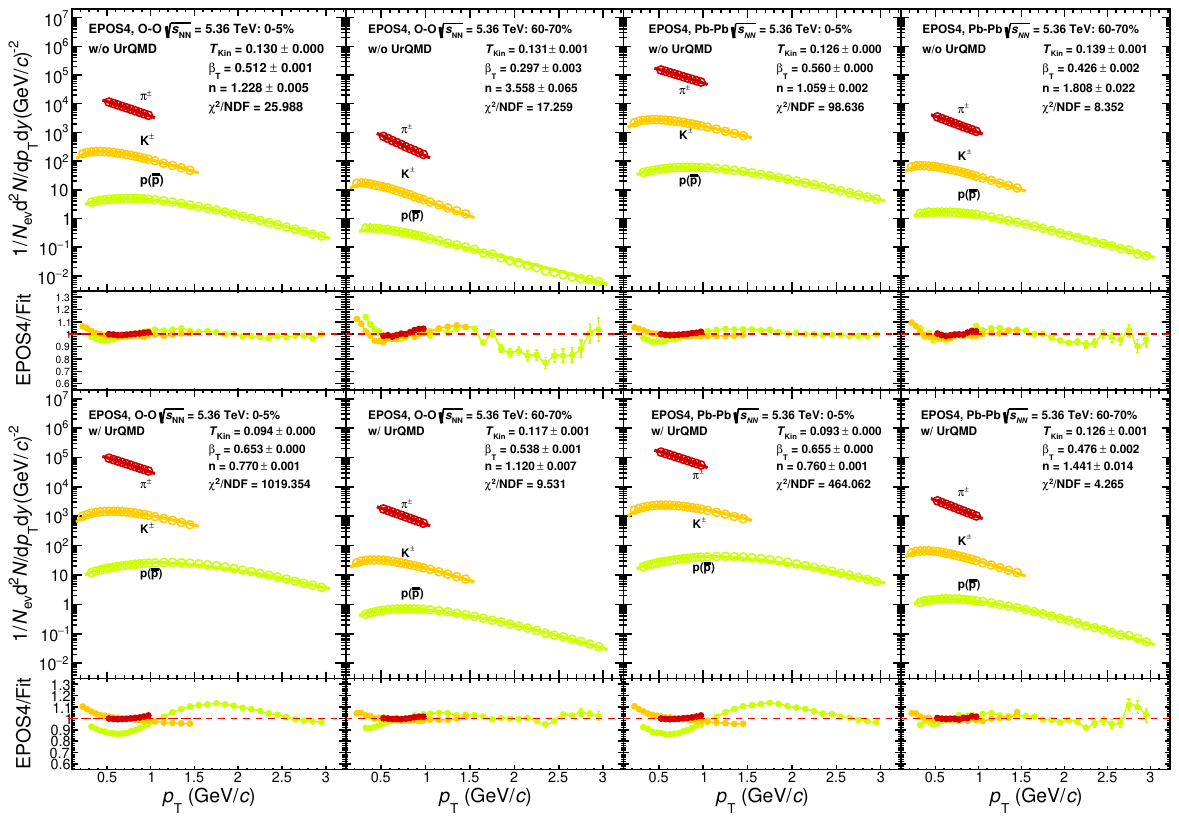}
    
    \caption{(Color online) (Top) Simultaneous Boltzmann–Gibbs blast-wave fits to identified particle spectra in O-O and Pb-Pb collisions at $\sqrt{s_{\text{NN}}} = 5.36$ TeV. Results are shown for the 0–5$\%$ and 60–70$\%$ centrality classes using event samples generated with EPOS4 with and without UrQMD. Solid lines represent the blast-wave fits to the corresponding spectra. (Bottom) Ratios of EPOS4 spectra to the blast-wave fits.}
    \label{BW_fit}
\end{figure}

\begin{figure}[h]
    \centering 
    \includegraphics[width=0.48\textwidth]{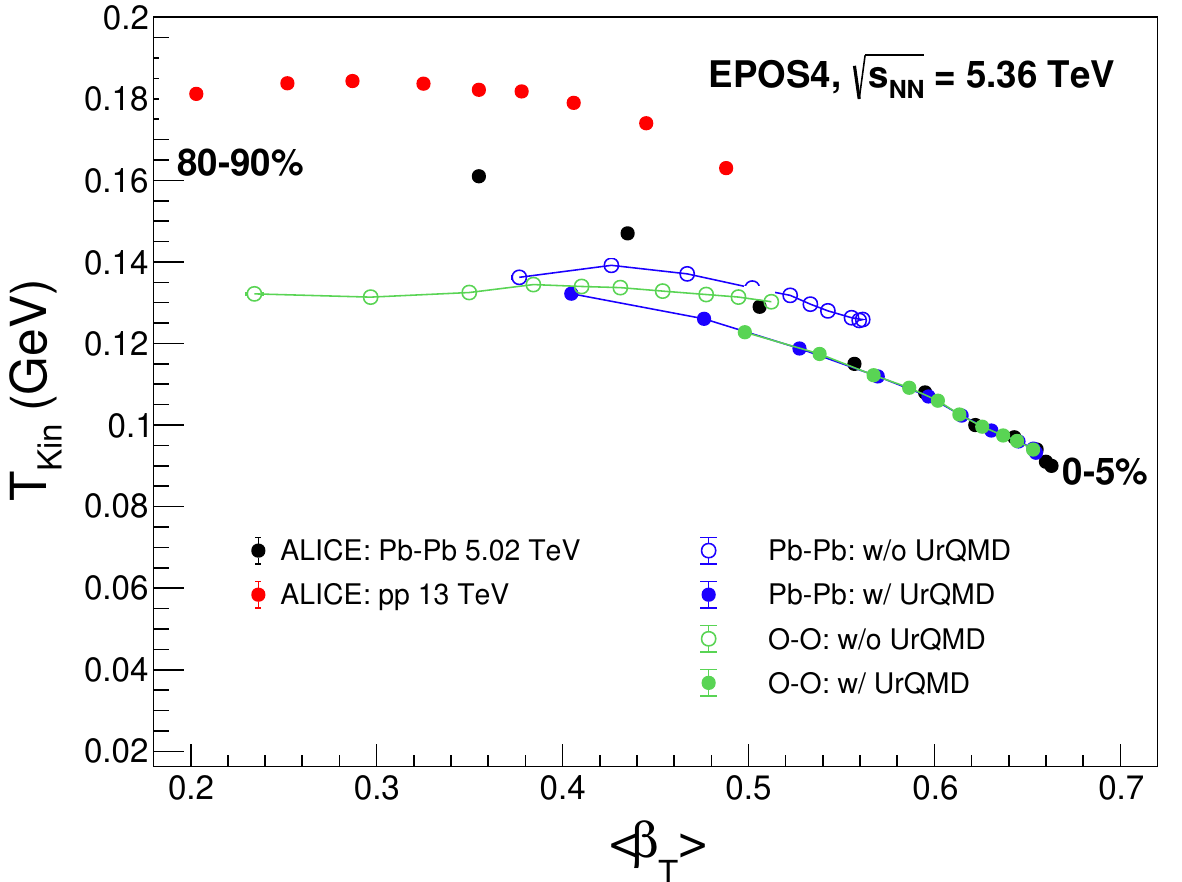}
    
    \caption{(Color online) Correlation between the kinetic freeze-out temperature $T_{\text{kin}}$ and the average transverse flow velocity $\langle \beta_T \rangle$ obtained from simultaneous Boltzmann–Gibbs blast-wave fits to $\pi^{\pm}$, $K^{\pm}$, $p$, $\overline{p}$ spectra in O–O and Pb–Pb collisions. The centrality increases from left to right. Results are shown for EPOS4 calculations with and without the UrQMD hadronic afterburner and compared with available ALICE Run 2 data \cite{Acharya_2020, PhysRevC.48.2462}.}
    \label{betaT_Combined}
\end{figure}

\begin{figure}
	\centering 
	\includegraphics[width=\textwidth, angle=0]{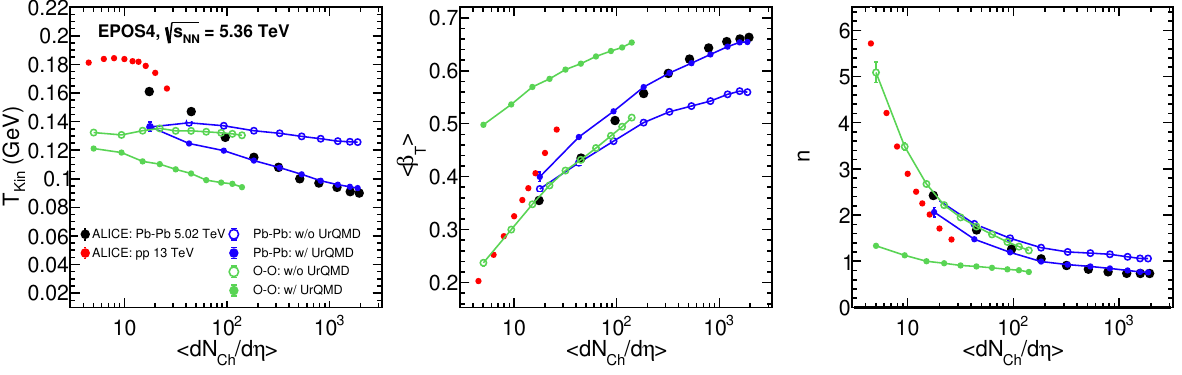}	
	\caption{(Color online) Multiplicity dependence of the kinetic freeze-out temperature $T_{\text{kin}}$, the average transverse flow velocity $\langle \beta_T \rangle$ and flow-profile exponent (n) obtained from simultaneous blast-wave fits to $\pi^{\pm}$, $K^{\pm}$, $p$, $\overline{p}$ spectra in O–O and Pb–Pb collisions at $\sqrt{s_{NN}}$ = 5.36 TeV. The fit ranges are 0.5–1 GeV/c for pions, 0.2–1.5 GeV/c for kaons, and 0.3–3 GeV/c for protons. Results are shown for EPOS4 calculations with and without the UrQMD hadronic afterburner and compared with available ALICE Run 2 data \cite{Acharya_2020, PhysRevC.48.2462}.} 
	\label{ParamwMult}%
\end{figure}

\subsection{Freeze-out Dynamics from Blast-Wave Analysis}\label{sec10}

The kinetic freeze-out stage fixes the final transverse-momentum spectra of hadrons and thus provides direct information on the interplay between thermal motion and collective expansion in high-energy nuclear collisions. In heavy-ion collisions, the system is generally expected to undergo two successive decoupling stages: chemical freeze-out, where inelastic reactions cease and the hadrochemical composition is fixed, followed by kinetic freeze-out, where elastic scatterings terminate and the momentum distributions no longer evolve. In contrast, pp collisions proceed on much shorter time scales and at lower energy densities, and are often described effectively by a single freeze-out scenario.

To quantify the kinetic freeze-out conditions and the strength of radial expansion, a simultaneous Boltzmann--Gibbs Blast-Wave (BG--BW) parameterization \cite{PhysRevC.48.2462} is applied to the transverse-momentum spectra of identified hadrons (\(\pi\), \(K\), and \(p\)). In this framework, particle emission is described as originating from a locally thermalized source undergoing collective transverse expansion, characterized by the kinetic freeze-out temperature \(T_{\mathrm{kin}}\) and the average transverse flow velocity \(\langle\beta_T\rangle\). The invariant yield is expressed as

\begin{equation}
    \frac{1}{p_{\rm{T}}} \frac{dN}{dp_{\rm{T}}} \propto \int_0^R r \, dr \, m_T I_0 \left( \frac{p_{\rm{T}} \sinh \rho}{T_{\text{kin}}} \right) K_1 \left( \frac{m_T \cosh \rho}{T_{\text{kin}}} \right),
\end{equation}
where \(m_T=\sqrt{p_T^2+m^2}\) is the transverse mass, \(I_0\) and \(K_1\) are modified Bessel functions, and the transverse-flow rapidity is defined as
\begin{equation}
    \rho = \tanh^{-1} \beta_T(r)= \tanh^{-1} \left( \left( \frac{r}{R} \right)^n \beta_s \right).
\end{equation}
Here, $R$ is the transverse radius of the fireball at freeze-out, $\beta_s$ denotes the surface velocity, and $n$ controls the flow-velocity profile.

The fit ranges follow those used in the corresponding ALICE analyses~\cite{PhysRevC.101.044907}: \(0.5<p_T<1\)~GeV/\(c\) for \(\pi^{\pm}\), \(0.2<p_T<1.5\)~GeV/\(c\) for \(K^{\pm}\), and \(0.3<p_T<3\)~GeV/\(c\) for \(p(\bar{p})\). Representative simultaneous fits and EPOS4-to-fit ratios are shown in Fig.~\ref{BW_fit} for the 0--5\% and 60--70\% centrality classes in O--O and Pb--Pb collisions at \(\sqrt{s_{NN}}=5.36\)~TeV, using EPOS4 calculations with and without the UrQMD hadronic afterburner. Within the selected fit intervals, the BG--BW parameterization provides a satisfactory description of the EPOS4 spectra, with deviations typically within 15--20\%.

Figure~\ref{betaT_Combined} summarizes the extracted freeze-out parameters in the
$T_{\mathrm{kin}}$--$\langle\beta_T\rangle$ plane. A clear anti-correlation is
observed: event classes characterized by larger transverse flow correspond to lower $T_{\mathrm{kin}}$, consistent with later kinetic decoupling in more strongly expanding systems.
For Pb--Pb collisions, the EPOS4 results at \(\sqrt{s_{NN}}=5.36\)~TeV follow the same global trend as the ALICE Run~2 reference at \(\sqrt{s_{NN}}=5.02\)~TeV \cite{PhysRevC.48.2462}, supporting the stability of the procedure. In peripheral Pb--Pb collisions, EPOS4 at 5.36~TeV yields slightly lower \(T_{\mathrm{kin}}\) and higher \(\langle\beta_T\rangle\) than the Run~2 extraction, consistent with a modest increase of collective expansion at the higher collision energy.
Notably, the O–O points do not interpolate smoothly between pp and Pb–Pb values in this plane; instead, when comparing similar centrality classes, they cluster closer to the Pb–Pb trend. This indicates that centrality selection and late-stage hadronic dynamics influence the extracted freeze-out parameters beyond a simple system-size scaling.

The multiplicity evolution of the blast-wave parameters is presented in Fig.~\ref{ParamwMult}. A systematic dependence on hadronic re-scattering is
observed: in both O--O and
Pb--Pb collisions, disabling UrQMD results in larger $T_{\mathrm{kin}}$ and
smaller $\langle\beta_T\rangle$, whereas enabling UrQMD shifts the parameters toward lower
$T_{\mathrm{kin}}$ and higher $\langle\beta_T\rangle$. This demonstrates that
late-stage hadronic interactions enhance the collective expansion component and
modify the kinetic decoupling conditions \cite{PhysRevLett.111.082302}. With UrQMD enabled, the extracted
parameters exhibit the experimentally observed centrality evolution, i.e. an
increase of $\langle\beta_T\rangle$ accompanied by a decrease of
$T_{\mathrm{kin}}$ towards central collisions.
In addition, the flow-profile exponent $n$ decreases from peripheral to central collisions, indicating a progressive modification of the radial-velocity profile. The inclusion of UrQMD shifts $n$ systematically to smaller values in central events, consistent with a more strongly developed collective expansion pattern in large and longer-lived systems.

\section{Summary and Discussions}\label{sec11}

In this work, we performed a comprehensive investigation of global observables and identified hadron production in pp, O--O, and Pb--Pb collisions at LHC Run~3 energies within the EPOS4 framework. A central element of this analysis is the dynamical core--corona separation implemented in EPOS4, which provides a unified description of particle production across systems of very different geometric size by disentangling contributions from a hydrodynamically evolving core and a string-fragmentation--dominated corona. Calculations were also performed both with and without the UrQMD hadronic afterburner, allowing a controlled assessment of late-stage hadronic effects. By combining charged-particle and transverse-energy densities, identified-hadron $p_T$ spectra, integrated yields, mean transverse momentum, particle ratios, nuclear modification factors, and kinetic freeze-out parameters, we examined how medium-like phenomena evolve with system size and multiplicity.

EPOS4 reproduces the overall trends of particle production across systems. The midrapidity charged-particle density $\langle dN_{\rm ch}/d\eta \rangle$ increases monotonically with multiplicity and provides a common scaling variable. 
While the midrapidity charged-particle density exhibits a monotonic rise with $\langle N_{\rm part}\rangle$ in both O--O and Pb--Pb collisions, a comparison at fixed participant fraction $\langle N_{\rm part}\rangle/2A$ reveals that O--O results approach and slightly exceed the Pb--Pb trend at large scaled participant fractions. This reflects the enhanced role of multiplicity fluctuations and geometric biases in lighter nuclei rather than anomalous entropy production. A consistent hierarchy is observed in transverse-energy production: the ratio $(dE_T/d\eta)/(dN_{\rm ch}/d\eta)$ is systematically larger in O--O than in Pb--Pb collisions at comparable participant scaling, indicating a harder effective production scale in the lighter system.
The transverse-momentum spectra show the expected multiplicity-dependent hardening. The most pronounced shape modifications occur at low and intermediate $p_T$ and increase with hadron mass. Central events display an enhancement at intermediate $p_T$, stronger for protons, consistent with sizable radial expansion, while peripheral events exhibit a relative suppression in the same region. The mean transverse momentum $\langle p_T\rangle$ increases with multiplicity for all species with clear mass ordering but exhibits a pronounced system dependence: at fixed $\langle dN_{\rm ch}/d\eta \rangle$, pp collisions yield the largest $\langle p_T\rangle$, followed by O--O and Pb--Pb. This demonstrates that $\langle p_T\rangle$ does not obey universal multiplicity scaling and retains sensitivity to the underlying production dynamics. The energy dependence of $p_T$-spectra between $\sqrt{s_{NN}}=5.02$ and 5.36~TeV in Pb--Pb collisions remains modest at low and intermediate $p_T$, with only moderate enhancement at high $p_T$.

These systematic trends can be consistently interpreted within the core--corona framework. The extracted core fractions increase monotonically with $\langle dN_{\rm ch}/d\eta \rangle$ in all systems, reflecting the growing importance of hydrodynamic emission at higher multiplicities. The correlated evolution of spectral shapes and core fractions indicates that multiplicity-dependent modifications arise primarily from changing relative weights of core and corona components. At low multiplicity, particle production is dominated by corona emission, while at high multiplicity—particularly in O--O and Pb--Pb collisions—the core contribution becomes dominant. Even high-multiplicity pp events exhibit a substantial core fraction, although the microscopic origin of collectivity differs from that in heavy-ion collisions.
The integrated yields follow the expected hadrochemical hierarchy $\pi > K > p$, with only weak energy dependence. The core--corona decomposition shows that the transition from corona-dominated to core-dominated production occurs at different multiplicity scales for different hadron species. Kaons, which carry strangeness, receive an enhanced contribution from the thermalized core at comparatively moderate multiplicities, reflecting the well-known enhancement of strange-quark production in dense matter. Protons, despite their larger mass, exhibit a comparatively delayed transition because they do not benefit from strangeness enhancement and continue to receive a sizable contribution from baryon pair production in the corona. This species-dependent transition scale highlights the combined influence of mass, baryon number, and strangeness content on hadron production dynamics.
Particle ratios provide complementary constraints on hadrochemistry and hadronization dynamics. The $K/\pi$ ratio shows only mild multiplicity dependence, whereas the $p/\pi$ ratio exhibits a pronounced enhancement at intermediate $p_T$ and a clear evolution with multiplicity and system size. While EPOS4 reproduces the qualitative behavior, it underestimates the magnitude of the intermediate-$p_T$ $p/\pi$ enhancement, indicating remaining tensions in the modelling of baryon production.
The multiplicity dependence of $p_T$-integrated ratios further constrains the dynamics. The integrated $p/\pi$ ratio exhibits a non-trivial evolution with multiplicity, whereas the integrated $K/\pi$ ratio shows a much weaker dependence and is described reasonably well within uncertainties. Comparisons of double ratios between Pb--Pb and O--O collisions at comparable $\langle N_{\rm part}\rangle$ indicate broad compatibility over most of the $p_T$ range, with modest deviations at low and intermediate $p_T$, consistent with differences in core and corona weights.
Medium-induced effects quantified via the nuclear modification factor $R_{AA}$ show strong suppression in Pb--Pb collisions that increases toward central events. EPOS4 predicts somewhat stronger suppression than available ALICE data at $\sqrt{s_{NN}}=5.02$~TeV in the intermediate-$p_T$ region. Notably, central O--O collisions at $\sqrt{s_{NN}}=5.36$~TeV also exhibit substantial suppression, demonstrating that medium-like effects persist in the intermediate system. The integrated $R_{AA}$ decreases continuously with $\langle N_{\rm part}\rangle$, with no indication of saturation in the explored range. The high-$p_T$ interval exhibits a steeper centrality dependence than the intermediate-$p_T$ region, reflecting the enhanced sensitivity of hard probes to path length and medium density.

The comparison of calculations with and without UrQMD allows a quantitative assessment of late-stage hadronic effects. Bulk observables are only weakly modified, whereas identified-hadron observables show measurable sensitivity. UrQMD leads to a moderate hardening of spectra and increases $\langle p_T\rangle$, most visibly for protons, consistent with baryon--antibaryon annihilation and re-scattering. The suppression of the integrated $p/\pi$ ratio in the highest-multiplicity Pb--Pb events is reproduced only when UrQMD is included, underscoring the importance of hadronic-phase interactions. Blast-wave fits confirm the characteristic anti-correlation between $T_{\rm kin}$ and $\langle\beta_T\rangle$, with hadronic re-scattering shifting the system toward lower kinetic temperatures and higher transverse flow velocities. The extracted flow-profile exponent $n$ exhibits a systematic evolution with centrality, indicating modifications of the radial-velocity profile. Together, these results demonstrate that late-stage hadronic re-scattering plays a non-negligible role in kinetic decoupling while leaving bulk observables largely unchanged.

In conclusion, EPOS4 provides a unified description of particle production across collision systems from pp to Pb--Pb at LHC Run~3 energies. The explicit core--corona separation, combined with controlled inclusion of hadronic re-scattering, shows that the observed trends emerge from the interplay between hydrodynamic collectivity, fragmentation-like emission, and late-stage hadronic dynamics. The O--O system serves as a crucial intermediate benchmark, revealing that similar final-state multiplicities can arise from qualitatively different microscopic mechanisms and providing a sensitive probe of the onset and evolution of medium-like effects.

\acknowledgments

The authors thank Prof. Klaus Werner for providing us with the EPOS4 model. The authors are also grateful to Dr. Subikash Choudhury for having fruitful discussion.
The authors are thankful to the members of the grid computing team of VECC for providing uninterrupted facilities for event generation and analyses. 
We also gratefully acknowledge the financial help from the DST-GOI under the scheme “Mega facilities for basic science research” [Sanction Order No. SR/MF/PS-02/2021-Jadavpur (E-37128) dated December 31, 2021].

\bibliographystyle{JHEP}
\bibliography{biblio.bib}

\end{document}